\documentclass[aps,prd,twocolumn,superscriptaddress]{revtex4}
\usepackage{epsfig,epsf}
\usepackage{amsmath}
\usepackage{amsthm}
\usepackage{amsfonts}
\usepackage{amssymb}
\usepackage{dsfont}
\usepackage{multirow}
\usepackage{appendix}
\usepackage{slashed}
\usepackage[active]{srcltx}
\usepackage{psfrag}

\begin{document}

\title{Resonance $X(4630)$}
\date{\today}
\author{S.~S.~Agaev}
\affiliation{Institute for Physical Problems, Baku State University, Az--1148 Baku,
Azerbaijan}
\author{K.~Azizi}
\affiliation{Department of Physics, University of Tehran, North Karegar Avenue, Tehran
14395-547, Iran}
\affiliation{Department of Physics, Do\v{g}u\c{s} University, Dudullu-\"{U}mraniye, 34775
Istanbul, Turkiye}
\author{H.~Sundu}
\affiliation{Department of Physics, Kocaeli University, 41380 Izmit, Turkiye}

\begin{abstract}
We investigate the structure $X(4630)$ discovered by the LHCb collaboration
in the process $B^{+}\rightarrow J/\psi \phi K^{+}$ as a resonance in the $%
J/\psi \phi $ mass distribution. We explore this resonance as a
diquark-antidiquark state $X=[cs][\overline{c}\overline{s}]$ with
spin-parities $J^{\mathrm{PC}}=1^{-+}$. Its mass and current coupling are
calculated using the QCD two-point sum rule method by taking into account
vacuum condensates up to dimension $10$. We also study decays of this
tetraquark to mesons $J/\psi\phi $, $\eta _{c}\eta ^{(\prime )}$, and $\chi
_{c1}\eta ^{(\prime )}$, and compute partial widths of these channels. To
this end, we employ the light-cone sum rule approach and technical methods
of soft-meson approximation to extract strong coupling at relevant
tetraquark-meson-meson vertices. Our predictions for the mass $m=(4632\pm
60)~\mathrm{MeV}$ and width $\Gamma=(159\pm 31)~\mathrm{MeV}$ of $X$ are in
a very nice agreement with recent measurements of the LHCb collaboration.
These results allow us to interpret the resonance $X(4630)$ as the
tetraquark $X$ with spin-parities $J^{\mathrm{PC}}=1^{-+}$.
\end{abstract}

\maketitle


\section{Introduction}

\label{sec:Int} 

Recently the LHCb collaboration informed about new charmonium-like
resonances $Z_{cs}$ and $X$ \ observed in the process $B^{+}\rightarrow
J/\psi \phi K^{+}$ in $J/\psi K^{+}$ and $J/\psi \phi $ invariant mass
distributions \cite{LHCb:2021uow}. The new resonances $Z_{cs}(4000)^{+}$ and
$Z_{cs}(4220)^{+}$ were discovered in the $J/\psi K^{+}$ channel, and are
presumably exotic mesons with a quark content $c\overline{c}u\overline{s}$.
States fixed in the $J/\psi \phi $ channel should be composed of $c\overline{%
c}s\overline{s}$ quarks provided they are four-quark structures. New
resonances in this channel $X(4630)$ and $X(4685)$ enriched a list of
vector, axial-vector and scalar states discovered by LHCb during the last few years
\cite{Aaij:2016iza,LHCb:2016nsl}. The collaboration also updated parameters
of states seen at early stages of investigations.

These experimental results generated a theoretical activity aimed to explain
obtained information in the context of various approaches of high energy
physics. Studies were concentrated mainly around the resonances $X(4630)$
and $Z_{cs}$, in which authors calculated masses and magnetic moments of these
states, and  explored their decay channels \cite%
{Liu:2021xje,Ozdem:2021yvo,Yang:2021sue,Wang:2021ghk,Turkan:2021ome}. Some
of new states were explained as threshold effects as well \cite{Ge:2021sdq}.

The structure $X(4630)$ is a wide resonance with the mass
\begin{equation}
m_{\exp }=(4626\pm 16_{-110}^{+18})~\mathrm{MeV},  \label{eq:MassExp}
\end{equation}%
and width
\begin{equation}
\Gamma _{\exp }=(174\pm 27_{-73}^{+134})~\mathrm{MeV},  \label{eq:WidthExp}
\end{equation}%
respectively. The LHCb determined also the spin-parity of $X(4630)$ and
fixed them $J^{\mathrm{P}}=1^{-}$.

It should be noted that, a vector structure $Y(4626)$ with the mass $%
4625.9_{-6.0}^{+6.2}(\mathrm{stat.})\pm 0.4(\mathrm{sys.})~\mathrm{MeV}$ and
the width $49.8_{-11.5}^{+13.9}(\mathrm{stat.})\pm 4.0(\mathrm{sys.})~%
\mathrm{MeV}$ was seen by the Belle collaboration recently in the process $%
e^{+}e^{-}\rightarrow D_{s}^{\ast }D_{s1}(2536)$ \cite{Belle:2019qoi}. This
resonance can be considered as a member of $Y$ family of vector states
discovered in electron-positron annihilations. Other members of this group
are resonances $Y(4630)$ and $Y(4660)$. First of them was detected by Belle
in the process $e^{+}e^{-}\rightarrow \Lambda _{c}^{+}\Lambda _{c}^{-}$ as a
peak in the $\Lambda _{c}^{+}\Lambda _{c}^{-}$ invariant mass distribution
\cite{Pakhlova:2008vn}. Its parameters $m=4634_{-7}^{+8}(\mathrm{stat.}%
)_{-8}^{+5}(\mathrm{sys.})~\mathrm{MeV}$ and $\Gamma =92_{-24}^{+40}(\mathrm{%
stat.})_{-21}^{+10}(\mathrm{sys.})~\mathrm{MeV}$ are close to ones of the
resonance $Y(4626)$, and whether they are different states or not is under
investigation. It is interesting that $Y(4630)$ was usually identified with the
vector state $Y(4660)$ \cite{Dai:2017fwx}.

The resonance $Y(4660)$, as a particle produced in $e^{+}e^{-}$
annihilation, bears the quantum numbers $J^{\mathrm{PC}}=1^{--}$. It was
modeled as excited $5{}^{3}S_{1}$ and $6{}^{3}S_{1}$ charmonia, as a
compound of the scalar $f_{0}(980)$ and vector $\psi (2S)$ mesons, or as a
baryonium state. In our work \cite{Sundu:2018toi}, we explored $Y(4660)$ by
treating it as the diquark-antidiquark vector state $[cs][\overline{c}%
\overline{s}]$. We calculated the mass and current coupling of the
tetraquark $[cs][\overline{c}\overline{s}]$, and also evaluated its full
width. Our results for the mass and full width of the state $[cs][\overline{c%
}\overline{s}]$ allowed us to interpret it as the observed resonance $%
Y(4660) $.

From analysis of the decay channel $X(4630)\rightarrow J/\psi \phi $, it is
clear, that $X(4630)$ is charmonium-like state probably with hidden strange
component $s\overline{s}$. Then, in the four-quark model its quark content \
should be $c\overline{c}s\overline{s}$. It is also evident that $C$-parity
conservation implies that $X(4630)$ is $C$ parity positive particle, i.e.,
the quantum numbers of this resonance should be $J^{\mathrm{PC}}=1^{-+}$. In
other words, it can be considered as $C=+1$ counterpart of the resonance $%
Y(4626)$. Spin-parities $J^{\mathrm{PC}}=1^{-+}$ exclude interpretation of $%
X(4630)$ as an ordinary meson, because these quantum numbers are not
accessible in the conventional quark-antiquark model. In other words, the
resonance $X(4630)$ may be  a double-exotic state: It is composed of four
quarks and carries exotic quantum numbers.

Four valence quarks can be grouped in different ways to form a single
structure. Indeed, they may form two conventional colorless mesons and
constitute a hadronic molecule. Alternatively, four quarks $c\overline{c}s%
\overline{s}$ may build a diquark-antidiquark state $[cs][\overline{c}%
\overline{s}]$. The resonance $X(4630)$ was examined in the context of both
these models. Thus, it was considered in Ref.\ \cite{Yang:2021sue} as the
molecule $D_{s}^{\ast }\overline{D}_{s1}(2536)$ with required spin-parities.
\ An analysis was performed there using the one-boson-exchange method. The
mass of the molecule $D_{s}^{\ast }\overline{D}_{s1}$ was found equal to $%
4644~\mathrm{MeV}$ which is consistent with the LHCb data. The authors also
emphasized, that a decay to a meson pair $J/\psi \phi $ is the main decay
channel of the molecule $D_{s}^{\ast }\overline{D}_{s1}$.

The molecule model for $Y(4626)$ was used in Ref.\ \cite{He:2019csk}, in
which it was examined as a system $J^{\mathrm{PC}}=1^{--}$ appearing from
the interaction $D_{s}^{\ast }\overline{D}_{s1}-D_{s}\overline{D}_{s1}$. In
this article structures with spin-parities $J^{\mathrm{PC}}=0^{--},\
0^{-+},\ 1^{-+}$ and others were explored as well. This treatment for the
masses of the molecules $D_{s}^{\ast }\overline{D}_{s1}$ with $J^{\mathrm{PC}%
}=1^{--}$ and $J^{\mathrm{PC}}=1^{-+}$ leads to predictions $4646~\mathrm{MeV%
}$ and $4648~\mathrm{MeV}$, respectively. Heavy-antiheavy hadronic molecules
built of the $S$-wave charmed mesons and baryons were studied also in Ref.\
\cite{Dong:2021juy}. The authors assumed that interaction between mesons
(baryons) is saturated by a meson exchange, and searched for poles in such
systems by solving the Bethe-Salpeter equation.

In the framework of the QCD sum rule method a diquark-antidiquark option was
considered in Ref.\ \cite{Wang:2013exa}. The result of this article for the
mass of the tetraquark $[cs][\overline{c}\overline{s}]$ with $J^{\mathrm{PC}%
}=1^{-+}$ equals to $4.63_{-0.08}^{+0.11}~\mathrm{GeV}$ and agrees with the
new LHCb data. As is seen, almost all models for $X(4630)$ and predictions
for its mass extracted using various methods within errors are consistent
with experimental data. Stated differently, masses of exotic states do not
provide information sufficient to verify different models by confronting
them with each another or/and experimental data. Therefore, besides
computations of the mass, there is a necessity to evaluate the full width of
$X(4630)$ as precise as possible.

In the present work, we are going to fulfill this program and calculate the
mass and width of the resonance $X(4630)$. We treat $X(4630)$ as
diquark-antidiquark vector state $X=[cs][\overline{c}\overline{s}]$ with
spin-parities $J^{\mathrm{PC}}=1^{-+}$. Investigations are performed in the
context of the QCD sum rule method \cite{Shifman:1978bx,Shifman:1978by},
which is one of powerful nonperturbative tools of high energy physics. It
allows one to compute parameters not only of conventional mesons and
baryons, but also of multiquark hadrons \cite%
{Agaev:2020zad,Albuquerque:2018jkn}.

The mass and current coupling of the tetraquark $X$ are calculated in the
framework of the QCD two-point sum rule approach. In these calculations, we
take into account various quark, gluon, and mixed vacuum condensates up to
dimension $10$. To investigate numerous decay channels of $X$, we use the
light-cone sum rule (LCSR) method \cite{Balitsky:1989ry}. Most of
tetraquarks are strong-interaction unstable particles and decay into two
conventional mesons. The resonance $X(4630)$ decays primarily to a pair of
mesons $J/\psi \phi $ which is experimentally confirmed fact. In the present
work, we study decays of the tetraquark $X$ not only to $J/\psi \phi $, but
also to $\eta _{c}\eta ^{(\prime )}$, and $\chi _{c1}\eta ^{(\prime )} $
mesons saturating by these five channels its full width. The process $%
X\rightarrow J/\psi \phi $ is the dominant decay channel of the tetraquark $%
X $, whereas remaining modes are subdominant ones, but their contributions
are important to evaluate the full width of $X$.

Partial widths of aforementioned decays are determined by strong couplings
at relevant vertices. For instance, in the case of the dominant decay this
is a strong coupling $G$ at the vertex $XJ/\psi \phi $. Calculation of the
strong coupling at the tetraquark and two mesons vertex $XJ/\psi \phi $ in
the LCSR method necessitates usage of complementary technical tools. A
reason is that $X$ is built of four valence quarks, and the light-cone
expansion of the relevant nonlocal correlator leads to expressions which
instead of distribution amplitudes of the $\phi $ meson depend on its local
matrix elements. To preserve the four-momentum at the vertex $XJ/\psi \phi $%
, in this situation one needs to impose additional kinematical restriction
on the momentum of the $\phi $ meson. Troubles encountered afterward can be
handled by including into analysis technical methods known as a soft-meson
approximation \cite{Ioffe:1983ju,Braun:1995}. The soft-meson approximation
was adapted for investigation of tetraquarks in Ref.\ \cite{Agaev:2016dev},
and applied to explore decays some of such particles (see, for example,
Ref.\ \cite{Agaev:2020zad}). In present article, strong couplings at
relevant vertices are computed by including into analysis nonperturbative
terms up to dimension $8$. The coupling $G$ receives a contribution also
from twist $4$ matrix element of the $\phi $ meson.

This article is organized in the following way: The mass and current
coupling of the tetraquark $X$ are computed in Section\ \ref{sec:Mass}. We
calculate the strong coupling $G$ of particles at the vertex $XJ/\psi \phi $
in Section\ \ref{sec:Decay1}. Here, we find also the partial width of the
decay $X\rightarrow J/\psi \phi $. Section\ \ref{sec:Decay2} is devoted to
analysis of the processes $X\rightarrow \eta _{c}\eta ^{(\prime )}$ and $%
X\rightarrow \chi _{c1}\eta ^{(\prime )}$, and to computation of their
partial widths. To this end, we calculate couplings $g_{1}$ and $g_{2}$
corresponding to vertices $X\eta _{c}\eta $ and $X\eta _{c}\eta ^{\prime }$,
respectively. Strong couplings $g_{3}$ and $g_{4}$ required to study decays $%
X\rightarrow \chi _{c1}\eta ^{(\prime )}$ are found also in this section. In
Sec. \ref{sec:Conclusions}, we confront our results with LHCb data for the
resonance $X(4630)$. This section contains also our concluding remarks.


\section{Mass and current coupling of the tetraquark $X$}

\label{sec:Mass}
Sum rules to calculate the mass $m$ and current coupling $f$ \ of the
tetraquark $X$ can be derived from analysis of the correlation function
\begin{equation}
\Pi _{\mu \nu }(p)=i\int d^{4}xe^{ipx}\langle 0|\mathcal{T}\{J_{\mu
}(x)J_{\nu }^{\dag }(0)\}|0\rangle ,  \label{eq:CorrF1}
\end{equation}%
where $J_{\mu }(x)$ is the interpolating current for the $X$ state, and $%
\mathcal{T}$ is time-ordered product of two currents.

The current with required properties has the following form
\begin{eqnarray}
J_{\mu }(x) &=&\epsilon \widetilde{\epsilon }\left[ s_{b}^{T}(x)C\gamma
_{5}c_{c}(x)\overline{s}_{d}(x)\gamma _{5}\gamma _{\mu }C\overline{c}%
_{e}^{T}(x)\right.  \notag \\
&&\left. -s_{b}^{T}(x)C\gamma _{\mu }\gamma _{5}c_{c}(x)\overline{s}%
_{d}(x)\gamma _{5}C\overline{c}_{e}^{T}(x)\right],  \label{eq:Curr1}
\end{eqnarray}%
where $\epsilon \widetilde{\epsilon }=\epsilon _{abc}\epsilon _{ade}$, and $%
a $, $b$, $c$, $d$, and $e$ are color indices. In expression above $C$ is
the charge conjugation matrix.

The current $J_{\mu }(x)$ describes the tetraquark composed of the color
antitriplet scalar diquark $\epsilon s^{T}C\gamma _{5}c$ (vector diquark $%
\epsilon s^{T}C\gamma _{\mu }\gamma _{5}c$) and color triplet vector
antidiquark $\widetilde{\epsilon }\overline{s}\gamma _{5}\gamma _{\mu }C%
\overline{c}^{T}$ (scalar antidiquark $\widetilde{\epsilon }\overline{s}%
\gamma _{5}C\overline{c}^{T}$). This current belongs to antitriplet-triplet
representation $\mathbf{[}\overline{\mathbf{3}_{\mathbf{c}}}\mathbf{]}%
_{cs}\otimes \mathbf{[3}_{\mathbf{c}}\mathbf{]}_{\overline{c}\overline{s}}$
of the color group $SU_{c}(3)$. Because the scalar diquark configuration is
most attractive and stable two-quark system \cite{Jaffe:2004ph}, the current
$J_{\mu }(x)$ corresponds to a ground-state vector particle with lowest mass
and required spin-parities.

To derive desired sum rules, we write down the correlation function $\Pi
_{\mu \nu }(q)$ using the mass and current coupling of the state $X$. For
these purposes, we insert into the correlation function a complete set of
states with quantum numbers of $X$ and carry out in Eq.\ (\ref{eq:CorrF1})
integration over $x$. As a result, we get
\begin{equation}
\Pi _{\mu \nu }^{\mathrm{Phys}}(p)=\frac{\langle 0|J_{\mu }|X(p,\varepsilon
\rangle \langle X(p,\varepsilon )|J_{\nu }^{\dagger }|0\rangle }{m^{2}-p^{2}}%
+\cdots ,  \label{eq:PhysSide}
\end{equation}%
with $m$ being the mass of $X$. In Eq.\ (\ref{eq:PhysSide}) dots stand for
contributions of higher resonances and continuum states. We introduce the
current coupling $f$ by means of the matrix element
\begin{equation}
\langle 0|J_{\mu }|X(p,\varepsilon )\rangle =fm\varepsilon _{\mu },
\label{eq:Res}
\end{equation}%
where $\varepsilon _{\mu }$ is the polarization vector of the tetraquark $X$%
. In terms of $m$ and $f$, the correlation function can be rewritten in the
following form%
\begin{equation}
\Pi _{\mu \nu }^{\mathrm{Phys}}(p)=\frac{m^{2}f^{2}}{m^{2}-p^{2}}\left(
-g_{\mu \nu }+\frac{p_{\mu }p_{\nu }}{m^{2}}\right) +\cdots .
\label{eq:CorM}
\end{equation}

We calculate the QCD side of the correlation function $\Pi _{\mu \nu }(p)$
using explicit expression of the current $J_{\mu }(x)$ and obtain $\Pi _{\mu
\nu }^{\mathrm{OPE}}(p)$ in terms of heavy and light quark propagators. Then
for $\Pi _{\mu \nu }^{\mathrm{OPE}}(p)$, we get the following formula
\begin{eqnarray}
&&\Pi _{\mu \nu }^{\mathrm{OPE}}(p)=i\int d^{4}xe^{ipx}\epsilon \widetilde{%
\epsilon }\epsilon ^{\prime }\widetilde{\epsilon }^{\prime }\left\{ \mathrm{%
Tr}\left[ \gamma _{5}\widetilde{S}_{s}^{bb^{\prime }}(x)\gamma _{5}\right.
\right.  \notag \\
&&\left. \times S_{c}^{cc^{\prime }}(x)\right] \mathrm{Tr}\left[ \gamma
_{5}\gamma _{\mu }\widetilde{S}_{c}^{ee^{\prime }}(-x)\gamma _{\nu }\gamma
_{5}S_{s}^{dd^{\prime }}(-x)\right]  \notag \\
&&-\mathrm{Tr}\left[ \gamma _{5}\gamma _{\mu }\widetilde{S}_{c}^{e^{\prime
}e}(-x)\gamma _{5}S_{s}^{d^{\prime }d}(-x)\right] \mathrm{Tr}\left[ \gamma
_{5}\gamma _{\nu }\widetilde{S}_{s}^{bb^{\prime }}(x)\right.  \notag \\
&&\times \left. \gamma _{5}S_{c}^{cc^{\prime }}(x)\right] -\mathrm{Tr}\left[
\gamma _{5}\widetilde{S}_{c}^{e^{\prime }e}(-x)\gamma _{\nu }\gamma
_{5}S_{s}^{d^{\prime }d}(-x)\right]  \notag \\
&&\times \mathrm{Tr}\left[ \gamma _{5}\widetilde{S}_{s}^{bb^{\prime
}}(x)\gamma _{\mu }\gamma _{5}S_{c}^{cc^{\prime }}(x)\right] +\mathrm{Tr}%
\left[ \gamma _{5}\gamma _{\nu }\widetilde{S}_{s}^{bb^{\prime }}(x)\right.
\notag \\
&&\left. \left. \times \gamma _{\mu }\gamma _{5}S_{c}^{cc^{\prime }}(x)
\right] \mathrm{Tr}\left[ \gamma _{5}\widetilde{S}_{c}^{e^{\prime
}e}(-x)\gamma _{5}S_{s}^{dd^{\prime }}(-x)\right] \right\} ,
\label{eq:CorrF2}
\end{eqnarray}%
where $\epsilon ^{\prime }\widetilde{\epsilon }^{\prime }=\epsilon
_{a^{\prime }b^{\prime }c^{\prime }}\epsilon _{a^{\prime }d^{\prime
}e^{\prime }}$. In Eq.\ (\ref{eq:CorrF2}) $S_{s}^{ab}(x)$ and $S_{c}^{ab}(x)$
are the $s$ and $c$-quark propagators, respectively. Their explicit
expressions are collected in Appendix. Here, we also use the notation
\begin{equation}
\widetilde{S}_{s(c)}(x)=CS_{s(c)}^{T}(x)C.
\end{equation}

To continue our analysis, we have to choose same structures both in $\Pi
_{\mu \nu }^{\mathrm{Phys}}(p)$ and $\Pi _{\mu \nu }^{\mathrm{OPE}}(p)$. For
our purposes, it is convenient to work with terms proportional to $-g_{\mu
\nu }$, i. e., with invariant amplitude $\Pi ^{\mathrm{Phys}%
}(p^{2})=m^{2}f^{2}/(m^{2}-p^{2})+\cdots $. This function receives
contributions only from spin-$1$ particles and are free of contaminations.
The amplitude $\Pi ^{\mathrm{Phys}}(p^{2})$ can be expressed by the
dispersion integral%
\begin{equation}
\Pi ^{\mathrm{Phys}}(p^{2})=\int_{4\mathcal{M}^{2}}^{\infty }\frac{\rho ^{%
\mathrm{Phys}}(s)ds}{s-p^{2}}+\cdots ,  \label{eq:DisRel}
\end{equation}%
where $\mathcal{M=}m_{c}+m_{s}$, and dots indicate subtraction terms
necessary to make the whole expression finite. The imaginary part of the
amplitude $\Pi ^{\mathrm{Phys}}(p^{2})$ constitutes the spectral density $%
\rho ^{\mathrm{Phys}}(s)$, which can be written down in the following form%
\begin{equation}
\rho ^{\mathrm{Phys}}(s)=\frac{1}{\pi }\mathrm{Im}\Pi ^{\mathrm{Phys}%
}(s)=m^{2}f^{2}\delta (s-m^{2})+\rho ^{\mathrm{h}}(s).  \label{eq:SDensity}
\end{equation}%
Here, contribution of the ground-state particle (the pole term) is separated
from one due to higher resonances and continuum states: the latter is
characterized by an unknown hadronic spectral density $\rho ^{\mathrm{h}}(s)$%
. It is not difficult to see that $\rho ^{\mathrm{Phys}}(s)$ substituted
into Eq.\ (\ref{eq:DisRel}) leads to the expression of the ground-state term
\begin{equation}
\Pi ^{\mathrm{Phys}}(p^{2})=\frac{m^{2}f^{2}}{m^{2}-p^{2}}+\int_{4\mathcal{M}%
^{2}}^{\infty }\frac{\rho ^{\mathrm{h}}(s)ds}{s-p^{2}}.  \label{eq:InvAmp2}
\end{equation}%
The obtained formula contains also a contribution coming from higher resonances
and continuum states.

The amplitude $\Pi ^{\mathrm{OPE}}(p^{2})$ can be calculated theoretically
in deep Euclidean region $p^{2}\ll 0$ in the operator product expansion ($%
\mathrm{OPE}$) with certain accuracy. The coefficient functions in this
expansion could be found using methods of perturbative QCD (PQCD), whereas
nonperturbative information is encoded by vacuum expectation values of
various quark, gluon and mixed operators. Having continued $\Pi ^{\mathrm{OPE%
}}(p^{2})$ analytically to the Minkowski domain and computed its imaginary
part one determines the two-point spectral density $\rho ^{\mathrm{OPE}}(s)$%
. In the region $p^{2}\ll 0$ one applies also the Borel transformation to
remove subtraction terms in the dispersion integral and suppress
contributions of higher resonances and continuum states. In the case of $\Pi
^{\mathrm{Phys}}(p^{2})$, we find
\begin{equation}
\mathcal{B}\Pi ^{\mathrm{Phys}}(p^{2})=m^{2}f^{2}e^{-m^{2}/M^{2}}+\int_{4%
\mathcal{M}^{2}}^{\infty }ds\rho ^{\mathrm{h}}(s)e^{-s/M^{2}},
\label{eq:CorBor}
\end{equation}%
with $M^{2}$ being the Borel parameter. Similar dispersion representation
can be written down for $\Pi ^{\mathrm{OPE}}(p^{2})$ in terms of $\rho ^{%
\mathrm{OPE}}(s)$ as well. Later, using assumption about hadron-parton
duality and matching $\rho ^{\mathrm{h}}(s)\simeq \rho ^{\mathrm{OPE}}(s)$
in duality region, it is possible to subtract second term in Eq.\ (\ref%
{eq:CorBor}) from the QCD side of the sum rule and get
\begin{equation}
m^{2}f^{2}e^{-m^{2}/M^{2}}=\int_{4\mathcal{M}^{2}}^{s_{0}}ds\rho ^{\mathrm{%
OPE}}(s)e^{-s/M^{2}}+\Pi (M^{2}),  \label{eq:SR}
\end{equation}%
where $s_{0}$ is continuum subtraction parameter. The second component of
the invariant amplitude $\Pi (M^{2})$ contains nonperturbative contributions
computed directly from $\Pi _{\mu \nu }^{\mathrm{OPE}}(p)$.

As is seen, physical parameters $m$ and $f$ of the tetraquark are expressed
in terms of $\rho ^{\mathrm{OPE}}(s)$ and $\Pi (M^{2})$ calculated in
quark-gluon degrees of freedom. To complete a system of equations and
determine the mass and coupling of the tetraquark $X$, we act by the
operator $d/d(-1/M^{2})$ to both sides of the equality Eq.\ (\ref{eq:SR}),
and, by this way, find missed second expression. This system can be solved,
and sum rules for the mass $m$ and coupling $f$ read
\begin{equation}
m^{2}=\frac{\Pi ^{\prime }(M^{2},s_{0})}{\Pi (M^{2},s_{0})},  \label{eq:Mass}
\end{equation}%
and%
\begin{equation}
f^{2}=\frac{e^{m^{2}/M^{2}}}{m^{2}}\Pi (M^{2},s_{0}).  \label{eq:Coupling}
\end{equation}%
Here, we denote r.h.s. of Eq.\ (\ref{eq:SR}) as $\Pi (M^{2},s_{0})$, and
introduce also a function $\Pi ^{\prime }(M^{2},s_{0})=d\Pi
(M^{2},s_{0})/d(-1/M^{2})$.

In the present article, $\Pi (M^{2},s_{0})$ is calculated at the leading
order of PQCD by taking into account quark, gluon and mixed vacuum
condensates up to dimension $10$. Details of computations of the spectral
density $\rho ^{\mathrm{OPE}}(s)$ and function $\Pi (M^{2})$ can be found,
for instance, in Ref.\ \cite{Agaev:2016dev}. Therefore, we do not consider
here these usual operations, and move explicit expression of the function $%
\Pi (M^{2},s_{0})$ to Appendix.

The sum rules for the mass and coupling given by Eqs.\ (\ref{eq:Mass}) and (%
\ref{eq:Coupling}) contain quark, gluon and mixed condensates which are
universal parameters of computations. They depend also masses of $c$ and $s$
quarks. Numerical values all of these parameters are listed below
\begin{eqnarray}
&&\langle \overline{q}q\rangle =-(0.24\pm 0.01)^{3}~\mathrm{GeV}^{3},\
\langle \overline{s}s\rangle =(0.8\pm 0.1)\langle \overline{q}q\rangle ,
\notag \\
&&\langle \overline{s}g_{s}\sigma Gs\rangle =m_{0}^{2}\langle \overline{s}%
s\rangle ,\ m_{0}^{2}=(0.8\pm 0.1)~\mathrm{GeV}^{2},\   \notag \\
&&\langle \frac{\alpha _{s}G^{2}}{\pi }\rangle =(0.012\pm 0.004)~\mathrm{GeV}%
^{4},  \notag \\
&&\langle g_{s}^{3}G^{3}\rangle =(0.57\pm 0.29)~\mathrm{GeV}^{6},  \notag \\
&&m_{c}=(1.275\pm 0.025)~\mathrm{GeV}.\ \ \ m_{s}=93_{-5}^{+11}~\mathrm{MeV}.%
\text{ }  \label{eq:Parameters}
\end{eqnarray}

The sum rules are functions also of auxiliary parameters $M^{2}$ and $s_{0}$%
, which have to obey standard constraints imposed on them by the sum rule
method. This means, that in the working regions of the parameters $M^{2}$
and $s_{0}$ a pole contribution ($\mathrm{PC}$) should dominate in the sum
rules and the operator product expansion should converge rapidly. To
quantify these constraints and use them to fix working windows for $M^{2}$
and $s_{0}$, we introduce expressions
\begin{equation}
\mathrm{PC}=\frac{\Pi (M^{2},s_{0})}{\Pi (M^{2},\infty )}.  \label{eq:Pole}
\end{equation}%
and
\begin{equation}
R(M^{2})=\frac{\Pi ^{\mathrm{DimN}}(M^{2},s_{0})}{\Pi (M^{2},s_{0})},
\label{eq:Convergence}
\end{equation}%
where $\Pi ^{\mathrm{DimN}}(M^{2},s_{0})$ is a contribution of the last
three terms in $\mathrm{OPE}$, i.e., $\mathrm{DimN=Dim(8+9+10)}$.

The formula Eq.\ (\ref{eq:Pole}) determines a contribution of the pole term
to the function $\Pi (M^{2},s_{0})$. In our present study, we adopt the
limit $\mathrm{PC\geqslant 0.2}$, which is typical for multiquark particles.
The convergence of the operator product expansion is examined by means of
the expression Eq.\ (\ref{eq:Convergence}): The convergence of $\mathrm{OPE}$
is fulfilled if at the minimum of the Borel parameter the ratio $R(M^{2})$
does not exceed $0.01$. The mass and current coupling of $X$ obtained by
means of the sum rules, in general, have not to depend on the Borel
parameter, but in actual computations, one can only limit its influence on
obtained predictions. Thus, a stability of extracted results is among
employed constraints to get the parameters $M^{2}$ and $s_{0}$.

Computations show that the working regions that meet all these constraints
are
\begin{equation}
M^{2}\in \lbrack 5.5,6.5]~\mathrm{GeV}^{2},\ s_{0}\in \lbrack 24,25]~\mathrm{%
GeV}^{2}.  \label{eq:Regions}
\end{equation}%
In fact, in these regions the pole contribution varies within a range $%
0.66\leq \mathrm{PC}\leq 0.26$. The convergence of $\mathrm{OPE}$ is also
satisfied, because at $M^{2}=5.5~\mathrm{GeV}^{2}$, we fix $R(M^{2})\leq
0.01 $.

\begin{widetext}

\begin{figure}[h!]
\begin{center}
\includegraphics[totalheight=6cm,width=8cm]{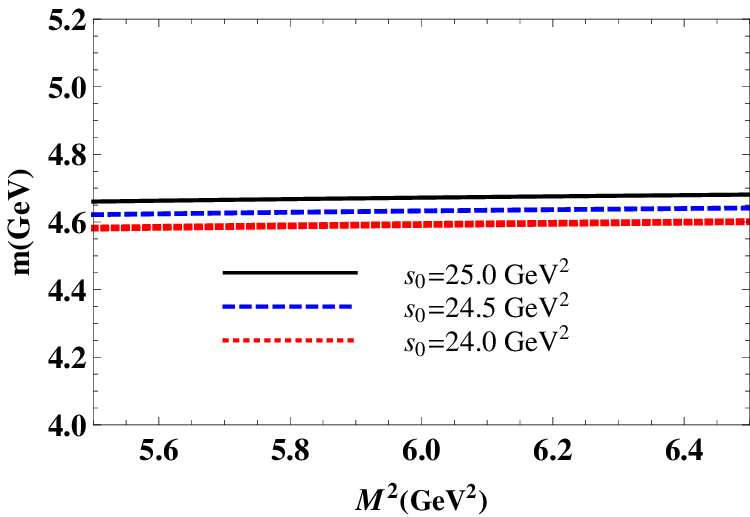}\,\, %
\includegraphics[totalheight=6cm,width=8cm]{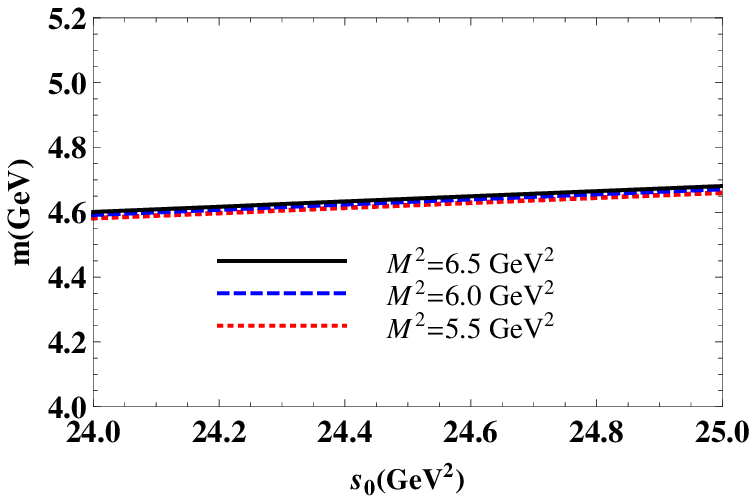}
\end{center}
\caption{ The mass of the tetraquark $X(4630)$ as a function of the
Borel parameter $M^2$ at fixed $s_0$ (left panel), and as a function
of the continuum threshold $s_0$ at fixed $M^2$ (right panel).}
\label{fig:Mass}
\end{figure}

\end{widetext}

To extract numerical values of the mass $m$ and coupling $f$, we calculate
them at different choices of the parameters $M^{2}$ and $s_{0}$, and find
their mean values averaged over the working regions Eq.\ (\ref{eq:Regions}).
For $m$ and $f$ these calculations yield
\begin{eqnarray}
m &=&(4632\pm 60)~\mathrm{MeV},  \notag \\
f &=&(9.2\pm 0.8)\times 10^{-3}~\mathrm{GeV}^{4}.  \label{eq:Result1}
\end{eqnarray}%
The values from Eq.\ (\ref{eq:Result1}) correspond to sum rules' results
computed at middle point of the working regions, i.e., to results at the
points $M^{2}=6~\mathrm{GeV}^{2}$ and $s_{0}=24.5~\mathrm{GeV}^{2}$. At this
point the pole contribution is $\mathrm{PC}\approx 0.51$, which guarantees
reliability of obtained predictions, and a ground-state nature of $X$.

In Fig.\ \ref{fig:Mass}, we plot the mass of the tetraquark $X$ as functions
of the parameters $M^{2}$ and $s_{0}$. As is seen, the mass $m$ is sensitive
to a choice of $M^{2}$ and $s_{0}$. It is also evident that within limits $%
M^{2}\in \lbrack 5.5,6.5]~\mathrm{GeV}^{2}$ this dependence is weak and
theoretical errors do not exceed $1.5\%$, whereas similar estimate for the
coupling gives $9\%$. This effect has simple explanation: The mass of the
tetraquark is determined by the ratio of the correlation functions Eq.\ (\ref%
{eq:Mass}). As a result, this ratio smooths dependence of $m$ on the
parameter $M^{2}$, which is not a case for the coupling Eq.\ (\ref%
{eq:Coupling}).

The mass of the tetraquark $X$ obtained in the present work is in excellent
agreement with the LHCb data for the mass of the resonance $X(4630)$. At
this phase of our studies, we can conclude that $X(4630)$ is the
diquark-antidiquark state $X=[cs][\overline{c}\overline{s}]$ with
spin-parities $J^{\mathrm{PC}}=1^{-+}$.


\section{ Decay $X\rightarrow J/\protect\psi \protect\phi $ \ }

\label{sec:Decay1}

The resonance $X(4630)$ was observed in the invariant mass distribution of
the $J/\psi \phi $ mesons. Hence, the process $X(4630)\rightarrow J/\psi
\phi $ can be considered as its dominant decay channel. In this section, we
consider this decay and calculate partial width of the process $X\rightarrow
J/\psi \phi $, which is governed by the strong coupling $G$ at the vertex $%
XJ/\psi \phi $.

In the context of the LCSR method the vertex\ $XJ/\psi \phi $ can be
explored by means of the correlator
\begin{equation}
\Pi _{\mu \nu }(p,q)=i\int d^{4}xe^{ipx}\langle \phi (q)|\mathcal{T}\{J_{\mu
}^{J/\psi }(x)J_{\nu }^{\dag }(0)\}|0\rangle ,  \label{eq:CorrF3}
\end{equation}%
with $J_{\nu }$ and $J_{\mu }^{J/\psi }$ being the interpolating currents of
the tetraquark $X$ and vector meson $J/\psi $, respectively. The $J_{\nu }$
is given by Eq.\ (\ref{eq:Curr1}), and current $J_{\mu }^{J/\psi }$ has the
form
\begin{equation}
J_{\mu }^{J/\psi }(x)=\overline{c}_{l}(x)\gamma _{\mu }c_{l}(x),
\label{eq:Bcur}
\end{equation}%
where $l=1,2,3$ is color index. In Eq.\ (\ref{eq:CorrF3}) $p$ and $q$ are
the momenta of the $J/\psi $ and $\phi $ mesons. Then the 4-momentum of the
the tetraquark $X$ is equal to $p^{\prime }=p+q$.

For on mass-shell $\phi $ meson $q^{2}=m_{\phi }^{2}$, the correlator $\Pi
_{\mu \nu }(p,q)$ is a function of two independent variables $p^{2}$ and $%
p^{\prime 2}=(p+q)^{2}$. It can be expanded over \ a set of Lorentz
structures in terms of invariant amplitudes $\Pi _{i}(p^{2},p^{\prime 2})$
and mass factors $C_{i}(\{m^{2}\})$. \ For our purposes, it is convenient to
expand $\Pi _{\mu \nu }(p,q)$ in the following basis
\begin{eqnarray}
&&\Pi _{\mu \nu }(p,q)=\Pi _{1}(p^{2},p^{\prime
2})C_{1}(\{m^{2}\})\varepsilon _{\mu }^{\ast }(q)p_{\nu }  \notag \\
&&+\Pi _{2}(p^{2},p^{\prime 2})C_{2}(\{m^{2}\})\varepsilon _{\nu }^{\ast
}(q)p_{\mu }++\Pi _{3}(p^{2},p^{\prime 2})  \notag \\
&&\times C_{3}(\{m^{2}\})\varepsilon ^{\ast }(q)\cdot pp_{\mu }p_{\nu }+\Pi
_{4}(p^{2},p^{\prime 2})  \notag \\
&&\times C_{4}(\{m^{2}\})\varepsilon ^{\ast }(q)\cdot pg_{\mu \nu }+\cdots ,
\label{eq:PhysN1}
\end{eqnarray}%
where $\varepsilon ^{\ast }(q)$ is polarization vector of the $\phi $ meson.
The factors $C_{i}(\{m^{2}\})$ depend on some combination of particles'
masses $\{m^{2}\}=\{m^{2},m_{1}^{2},m_{\phi }^{2}\}$ , with $m_{1}$ and $%
m_{\phi }$ being masses of the $J/\psi $ and $\phi $ mesons, respectively.

The phenomenological side of the sum rule can be obtained from Eq.\ (\ref%
{eq:CorrF3}) by expressing $\Pi _{\mu \nu }(p,q)$ in terms of physical
parameters of particles involved into the decay process. To explain this
procedure, as an example, let us consider the amplitude $\Pi
_{1}(p^{2},p^{\prime 2})$. Using the double dispersion relation \cite%
{Braun:1995,Colangelo:1997rp}, for $\Pi _{1}(p^{2},p^{\prime 2})$ we get
\begin{eqnarray}
&&\Pi _{1}(p^{2},p^{\prime 2})=\int \int \frac{\rho _{1}^{\mathrm{h}%
}(s_{1},s_{2})ds_{1}ds_{2}}{(s_{1}-p^{\prime 2})(s_{2}-p^{2})}  \notag \\
&&+\int \frac{\rho _{11}^{\mathrm{h}}(s_{1})ds_{1}}{(s_{1}-p^{\prime 2})}%
+\int \frac{\rho _{21}^{\mathrm{h}}(s_{2})ds_{2}}{(s_{2}-p^{2})}.
\label{eq:PhysN1a}
\end{eqnarray}%
As is seen, Eq.\ (\ref{eq:PhysN1a}) contains also single dispersion
integrals which are necessary to make finite the whole expression.

The amplitude $\Pi _{1}(p^{2},p^{\prime 2})$ receives contributions from two
channels: First channel contains vector tetraquarks $[cs][\overline{c}%
\overline{s}]$, whereas second one is a channel of vector charmonia.
Separating in spectral density $\rho _{1}^{\mathrm{h}}(s_{1},s_{2})$
contributions of ground-state particles in these channels, i. e.,
contribution of the tetraquark $X$ and $J/\psi $ from effects of higher
resonances and continuum states, we can model $\rho _{1}^{\mathrm{h}%
}(s_{1},s_{2})$ in the form \cite{Colangelo:1997rp}
\begin{eqnarray}
&&\rho _{1}^{\mathrm{h}}(s_{1},s_{2})=Gfmf_{1}m_{1}\delta
(s_{1}-m^{2})\delta (s_{2}-m_{1}^{2})  \notag \\
&&+\rho _{1}^{\mathrm{h}}(s_{1},s_{2})\theta (s_{1}-s_{0})\theta
(s_{2}-s_{0}^{\prime }),  \label{eq:PhysN1b}
\end{eqnarray}%
where $G$ is the strong coupling, which should be extracted from relevant
sum rule. The doubly spectral density $\rho _{1}^{\mathrm{h}}(s_{1},s_{2})$
contains also the current coupling $f$ of the tetraquark $X$ and decay
constant $f_{1}$ of the $J/\psi $ meson, which are defined by Eq.\ (\ref%
{eq:Res}) and by the matrix element
\begin{equation}
\langle 0|J_{\mu }^{J/\psi }|J/\psi \left( p\right) \rangle
=f_{1}m_{1}\varepsilon _{\mu }(p),  \label{eq:DC2}
\end{equation}%
respectively. Here, $\varepsilon _{\mu }(p)$ is the polarization vector of
the $J/\psi $ meson.

Substituting $\rho _{1}^{\mathrm{h}}(s_{1},s_{2})$ into Eq.\ (\ref%
{eq:PhysN1a}), we find
\begin{eqnarray}
&&\Pi _{1}(p^{2},p^{\prime 2})=\frac{Gfmf_{1}m_{1}}{\left( p^{\prime
2}-m^{2}\right) \left( p^{2}-m_{1}^{2}\right) }C_{1}(\{m^{2}\})  \notag \\
&&+\underset{\sum }{\int \int }\frac{\rho _{1}^{\mathrm{h}%
}(s_{1},s_{2})ds_{1}ds_{2}}{(s_{1}-p^{\prime 2})(s_{2}-p^{2})}+\cdots ,
\label{eq:PhysN1c}
\end{eqnarray}%
where $\sum $ is a domain in the $(s_{1},s_{2})$ plane boundaries of which $%
(s_{0},\ s_{0}^{\prime })$ depend on parameters of a process under analysis.
For the sake of brevity, we do not write down here single dispersion
integrals and denote them by dots. The similar dispersion relations can be
written down for remaining amplitudes, as well. Because the strong coupling $%
G$ is the same for all structures \cite{Bracco:2007sg}, one gets
\begin{eqnarray}
&&\Pi _{\mu \nu }^{\mathrm{Phys}}(p,q)=\frac{Gfmf_{1}m_{1}}{\left( p^{\prime
2}-m^{2}\right) \left( p^{2}-m_{1}^{2}\right) }  \notag \\
&&\times \left[ C_{1}(\{m^{2}\})\varepsilon _{\mu }^{\ast }(q)p_{\nu
}+C_{2}(\{m^{2}\})\varepsilon _{\nu }^{\ast }(q)p_{\mu }\right.  \notag \\
&&{}+C_{3}(\{m^{2}\})\varepsilon ^{\ast }(q)\cdot pp_{\mu }p_{\nu
}+C_{4}(\{m^{2}\})  \notag \\
&&\left. \times \varepsilon ^{\ast }(q)\cdot pg_{\mu \nu }+\cdots \right]
+\Pi _{\mu \nu }^{(\mathrm{HR,C)}}(p,q).  \label{eq:PhysN2}
\end{eqnarray}%
Contributions stemming from higher resonances and continuum states are
denoted in Eq.\ (\ref{eq:PhysN2}) by $\Pi _{\mu \nu }^{(\mathrm{HR,C)}}(p,q)$%
. We are interested in detailed analysis of the first term in $\Pi _{\mu \nu
}^{\mathrm{Phys}}(p,q)$ \cite{Braun:1995}, with poles at $p^{2}$ and $%
p^{\prime 2}=(p+q)^{2}$.

The correlation function $\Pi _{\mu \nu }^{\mathrm{Phys}}(p,q)$ can be
written down in the factorized form
\begin{eqnarray}
&&\Pi _{\mu \nu }^{\mathrm{Phys}}(p,q)=\langle \phi (q)J/\psi \left(
p\right) |X(p^{\prime })\rangle \frac{\langle X(p^{\prime })|J_{\nu }^{\dag
}|0\rangle }{\left( p^{\prime 2}-m^{2}\right) }  \notag \\
&&\times \frac{\langle 0|J_{\mu }^{J/\psi }|J/\psi \left( p\right) \rangle }{%
\left( p^{2}-m_{1}^{2}\right) }+\cdots ,  \label{eq:PhysN}
\end{eqnarray}%
where $mf$ and $m_{1}f_{1}$ are replaced by relevant martix elements (up to
polararization vectors), whereas on-mass-shell matrix element $\langle \phi
(q)J/\psi \left( p\right) |X(p^{\prime })\rangle $ \ defines the strong
coupling $G\ $at the vertex $XJ/\psi \phi $. It can be modeled in the
following form
\begin{eqnarray}
&&\langle \phi (q)J/\psi \left( p\right) |X(p^{\prime })\rangle =G\left[
(q-p)_{\gamma }g_{\alpha \beta }-(p^{\prime }+q)_{\alpha }g_{\gamma \beta
}\right.  \notag \\
&&\left. +(p^{\prime }+p)_{\beta }g_{\gamma \alpha }\right] \varepsilon
^{\gamma }(p^{\prime })\varepsilon ^{\ast \alpha }(p)\varepsilon ^{\ast
\beta }(q).  \label{eq:Mel}
\end{eqnarray}%
Then from Eq.\ (\ref{eq:PhysN}) one can easily find that
\begin{eqnarray}
&&\Pi _{\mu \nu }^{\mathrm{Phys}}(p,q)=\frac{Gfmf_{1}m_{1}}{\left( p^{\prime
2}-m^{2}\right) \left( p^{2}-m_{1}^{2}\right) }  \notag \\
&&\times \left[ \frac{m_{1}^{2}-m^{2}-m_{\phi }^{2}}{m^{2}}\varepsilon _{\mu
}^{\ast }(q)p_{\nu }+\frac{m^{2}-m_{1}^{2}-m_{\phi }^{2}}{m_{1}^{2}}%
\varepsilon _{\nu }^{\ast }(q)p_{\mu }\right.  \notag \\
&&{}\left. -\frac{m^{2}+m_{1}^{2}-m_{\phi }^{2}}{m^{2}m_{1}^{2}}\varepsilon
^{\ast }(q)\cdot pp_{\mu }p_{\nu }+2\varepsilon ^{\ast }(q)\cdot pg_{\mu \nu
}+\cdots \right]  \notag \\
&&+\Pi _{\mu \nu }^{(\mathrm{HR,C)}}(p,q),  \label{eq:CorrF5}
\end{eqnarray}%
where ellipses inside of the square brackets stand for terms that vanish in
the limit $p^{\prime }\rightarrow p$ (see an explanation below). Comparing
the correlation function $\Pi _{\mu \nu }^{\mathrm{Phys}}(p,q)$ in Eq.\ (\ref%
{eq:CorrF5}) with one from Eq.\ (\ref{eq:PhysN2}), one sees that they
coincide with each other provided functions $C_{i}(\{m^{2}\})$ are given by
formulas%
\begin{eqnarray}
C_{1}(\{m^{2}\}) &=&\frac{m_{1}^{2}-m^{2}-m_{\phi }^{2}}{m^{2}},  \notag \\
C_{2}(\{m^{2}\}) &=&\frac{m^{2}-m_{1}^{2}-m_{\phi }^{2}}{m_{1}^{2}},  \notag
\\
C_{3}(\{m^{2}\}) &=&-\frac{m^{2}+m_{1}^{2}-m_{\phi }^{2}}{m^{2}m_{1}^{2}},
\notag \\
C_{4}(\{m^{2}\}) &=&2.
\end{eqnarray}

There are a few Lorentz structures in Eq.\ (\ref{eq:CorrF5}), which may be
employed to construct a sum rule equality. In the present work, we choose to
work with the structure $\sim \varepsilon _{\mu }^{\ast }(q)p_{\nu }$ and
denote relevant invariant amplitude by $\Pi ^{\mathrm{Phys}}(p^{2},p^{\prime
2})$.

At the next phase of studies, we have to calculate the correlation function $%
\Pi _{\mu \nu }^{\mathrm{OPE}}(p,q)$ using quark-gluon degrees of freedom.
To this end, we insert expressions of the currents $J_{\mu }^{J/\psi }(x)$
and $J_{\nu }^{\dag }(0)$ into Eq.\ (\ref{eq:CorrF3}), contract relevant
quark fields and replace them by corresponding quark propagators. In full
LCSR treatment of vertices, for instance, composed of three conventional
mesons, a final expression obtained for $\Pi _{\mu \nu }^{\mathrm{OPE}}(p,q)$
depends on propagators and distribution amplitudes (DAs) of a meson.
Afterwards, separating in the correlation function a chosen Lorentz
structure and corresponding invariant amplitude $\Pi ^{\mathrm{OPE}%
}(p^{2},(p+q)^{2})$, one should calculate it in the regions $%
s_{1}=(p+q)^{2}\ll 0$ and $s_{2}=p^{2}\ll 0$, where methods of PQCD are
applicable. After analytical continuation of $\Pi ^{\mathrm{OPE}%
}(s_{1},s_{2})$ to Minkowski domain, computation its imaginary part over
variables $s_{1}$ and $s_{2}$, one can determine a spectral density $\rho ^{%
\mathrm{OPE}}(s_{1},s_{2})$. Then using parton-hadron duality assumption $%
\rho ^{\mathrm{h}}(s_{1},s_{2})\simeq \rho ^{\mathrm{OPE}}(s_{1},s_{2})$ and
performing double Borel transformations over variables $p^{2}\ll 0$ and $%
p^{\prime 2}\ll 0$ to suppress effects of higher resonances and remove
single dispersion integrals, one finds a sum rule which expresses
an on-mass-shell three-meson coupling in terms of $\rho ^{\mathrm{OPE}%
}(s_{1},s_{2})$.

In the case under discussion, i. e., for tetraquark-meson-meson vertex $%
XJ/\psi \phi $ full LCSR scheme outlined above has to be modified. Reasons
for that are connected with features of the function $\Pi _{\mu \nu }^{%
\mathrm{OPE}}(p,q)$. In fact, QCD expression for $\Pi _{\mu \nu }^{\mathrm{%
OPE}}(p,q)$ obtained using quark propagators is given by the formula
\begin{eqnarray}
&&\Pi _{\mu \nu }^{\mathrm{OPE}}(p,q)=-i\int d^{4}xe^{ipx}\epsilon
\widetilde{\epsilon }\left\{ \left[ \gamma _{5}\widetilde{S}%
_{c}^{lc}(x)\gamma _{\mu }\right. \right.  \notag \\
&&\left. \left. \times \widetilde{S}_{c}^{el}(-x)\gamma _{\nu }\gamma _{5}%
\right] +\left[ \gamma _{\nu }\gamma _{5}\widetilde{S}_{c}^{lc}(x){}\gamma
_{\mu }\widetilde{S}_{c}^{el}(-x){}\gamma _{5}\right] \right\} _{\alpha
\beta }  \notag \\
&&\times \langle \phi (q)|\overline{s}_{\alpha }^{b}(0)s_{\beta
}^{d}(0)|0\rangle ,  \label{TranCF1}
\end{eqnarray}%
where $\alpha $ and $\beta $ are spinor indices.

As is seen, the function $\Pi _{\mu \nu }^{\mathrm{OPE}}(p,q)$ instead of $%
\phi $ meson's distribution amplitudes depends on its local matrix elements.
The emerged situation has simple explanation: The meson $J/\psi $ is composed of
$c$ quark and antiquark at $x$ which can be contracted only with $c$%
-antiquark and quark from the tetraquark $X$. As a result, remaining $s$%
-quark fields in the current $J_{\nu }^{\dag }(0)$ located at the space-time
position $x=0$ establish local matrix elements of the $\phi $ meson.

To understand consequences of this situation, it is convenient to perform
following transformations
\begin{equation}
\overline{s}_{\alpha }^{b}s_{\beta }^{d}\rightarrow \frac{1}{12}\delta
^{bd}\Gamma _{\beta \alpha }^{j}\left( \overline{s}\Gamma ^{j}s\right) ,
\label{eq:MatEx}
\end{equation}%
where $\Gamma ^{j}$ is the full set of Dirac matrices,
\begin{equation}
\Gamma ^{j}=\mathbf{1,\ }\gamma _{5},\ \gamma _{\mu },\ i\gamma _{5}\gamma
_{\mu },\ \sigma _{\mu \nu }/\sqrt{2}.
\end{equation}%
Let us note, that in Eq.\ (\ref{eq:MatEx}), we use also the projector onto a
color-singlet state $\delta ^{bd}/3$.

After these manipulations, it is easy to carry out a color summation. Later,
we substitute quark propagators into obtained expression and perform $4$%
-dimensional integration over $x$. This integration creates in the integrand
the delta function $\delta ^{4}(p^{\prime }-p)$, which as an argument
contains only four-momenta of the tetraquark $X$ and meson $J/\psi $.
Therefore, subsequent integration over $p$ or $p^{\prime }$ \ sets $%
p=p^{\prime }$, which is the consequence of the four-momentum conservation
at the vertex $XJ/\psi \phi $. Stated differently, to preserve the
four-momentum at the tetraquark-meson-meson vertex one has to choose $q=0$.
In the full LCSR this is known as the soft-meson approximation \cite%
{Braun:1995}. At vertices of ordinary mesons $q\neq 0$, and only in the
soft-meson limit, one equates $q$ to zero, whereas the
tetraquark-meson-meson vertex can be explored in the framework of the LCSR
method only for $q=0$. It is worth emphasizing that
tetraquark-tetraquark-meson vertices can be explored using the full LCSR
method: Correlation function of a such vertex depends on distribution
amplitudes of a final meson \cite{Agaev:2016srl,Sundu:2017xct,Agaev:2019coa}%
. For our purposes, it is important that both the soft-meson approximation
and full LCSR treatment of the ordinary mesons' vertices lead for the strong
couplings to very close numerical predictions \cite{Braun:1995}, hence our
treatment of the coupling $G$ should give a reliable result.

Equation (\ref{eq:MatEx}) applied to $\Pi _{\mu \nu }^{\mathrm{OPE}}(p,0)$
generates different local matrix elements of the $\phi $ meson, which are
known and can be used to find analytical expression and carry out numerical
computations. Analysis confirms, that only two matrix elements of the $\phi $
meson contribute to the correlation function. First of them is twist-$2$
matrix element
\begin{equation}
\langle \phi (q)|\overline{s}(0)\gamma _{\mu }s(0)|0\rangle =f_{\phi
}m_{\phi }\varepsilon _{\mu }^{\ast }(q),
\end{equation}%
where $f_{\phi }$ is the decay constant of $\phi $ meson. Second matrix
element, which survives in the soft-meson limit, has twist $4$ and is given
by the expression%
\begin{equation}
\langle \phi (q)|\overline{s}(0)g\widetilde{G}_{\mu \nu }\gamma ^{\nu
}\gamma _{5}s(0)|0\rangle =f_{\phi }m_{\phi }^{3}\zeta _{4\phi }\varepsilon
_{\mu }^{\ast }(q).
\end{equation}%
Here, $\widetilde{G}_{\mu \nu }=1/2\varepsilon _{\mu \nu \alpha \beta
}G^{\alpha \beta }$ is the gluon dual field-strength tensor. The parameter $%
\zeta _{4\phi }=\pm 0.02$ was determined from the sum rule analysis in Ref.\
\cite{Ball:2007zt}, and is small.

But before deriving the sum rule for the strong coupling $G$, the soft limit
should be implemented also in the physical expression of the correlation
function $\Pi _{\mu \nu }^{\mathrm{Phys}}(p,q)$. In the limit $q\rightarrow
0 $, the ground-state term in $\Pi _{\mu \nu }^{\mathrm{Phys}}(p,0)$ can be
modified with some accuracy in the following way
\begin{equation}
\frac{1}{\left( p^{\prime 2}-m^{2}\right) \left( p^{2}-m_{1}^{2}\right) }%
\rightarrow \frac{1}{\left( p^{2}-\widetilde{m}^{2}\right) ^{2}},
\end{equation}%
where $\widetilde{m}^{2}$ is equal to $(m^{2}+m_{1}^{2})/2$. After this
transformation instead of two single poles at $p^{\prime 2}=m^{2}$ and $%
p^{2}=m_{1}^{2}$, the function $\Pi ^{\mathrm{Phys}}(p^{2},0)$ acquires one
double-pole at $p^{2}=\widetilde{m}^{2}$.

Having fixed in $\Pi _{\mu \nu }^{\mathrm{OPE}}(p,0)$ an amplitude $\Pi ^{%
\mathrm{OPE}}(p^{2})$ which corresponds to the structure $\sim \varepsilon
_{\mu }^{\ast }(q)p_{\nu }$, and carried out calculations in the region $%
p^{2}\ll 0$ we find finally the spectral density $\rho ^{\mathrm{OPE}}(s)$.
But in the soft approximation the Borel transformation and subtraction
procedure require more careful considerations than in full LCSR treatment.
In the soft limit one performs Borel transformation over one variable $%
p^{2}\ll 0$, and in this case single dispersion integrals also contribute to
hadronic part of the sum rules. These non-vanishing contributions correspond
to transitions from the excited states in $X$ channel \cite{Braun:1995}.
Therefore, before carrying out the continuum subtraction they should be
excluded from $\mathcal{B}\Pi ^{\mathrm{Phys}}(p^{2})$ by means of some
prescription. This problem is solved by the operator \cite%
{Ioffe:1983ju,Braun:1995}
\begin{equation}
\mathcal{P}(M^{2},\widetilde{m}^{2})=\left( 1-M^{2}\frac{d}{dM^{2}}\right)
M^{2}e^{\widetilde{m}^{2}/M^{2}},
\end{equation}%
that acts to both sides of the sum rule. It eliminates unsuppressed terms in
the physical side, but modifies also QCD side of the sum rule. Then
contributions of higher resonances with regular behavior can be subtracted
from the QCD side using the quark-hadron duality assumption.

The sum rule for the strong coupling $G$ reads%
\begin{equation}
G=\frac{m}{fm_{1}f_{1}}\frac{\mathcal{P}(M^{2},\widetilde{m}^{2})\mathcal{B}%
\Pi ^{\mathrm{OPE}}(p^{2})}{m_{1}^{2}-m^{2}-m_{\phi }^{2}}.  \label{eq:SR1}
\end{equation}%
The Borel transformed and subtracted correlation function $\mathcal{B}\Pi ^{%
\mathrm{OPE}}(p^{2})$ has the following form
\begin{equation}
\mathcal{B}\Pi ^{\mathrm{OPE}}(p^{2})=\int_{4\mathcal{M}^{2}}^{s_{0}}ds\rho
^{\mathrm{pert.}}(s)e^{-s/M^{2}}+\overline{\Pi }(M^{2}).  \label{eq:DecayPer}
\end{equation}%
The integral in Eq.\ (\ref{eq:DecayPer}) is a perturbative term, where the
spectral density $\rho ^{\mathrm{pert.}}(s)$ is determined by the expression
\begin{equation}
\rho ^{\mathrm{pert.}}(s)=\frac{f_{\phi }m_{\phi }m_{c}}{4\pi ^{2}}\frac{%
\sqrt{s(s-4m_{c}^{2})}}{s}.  \label{eq:SDen}
\end{equation}%
The second component of $\mathcal{B}\Pi ^{\mathrm{OPE}}(p^{2})$, i.e., the
function $\overline{\Pi }(M^{2})$ contains the twist-4 and nonperturbative
contributions,
\begin{eqnarray}
\overline{\Pi }(M^{2}) &=&\frac{f_{\phi }m_{\phi }^{3}m_{c}\zeta _{4\phi }}{%
16\pi ^{2}}\int_{0}^{1}\frac{dx}{x(x-1)}e^{-m_{c}^{2}/M^{2}x(1-x)}  \notag \\
&&+\frac{f_{\phi }m_{\phi }m_{c}}{4}\mathcal{F}^{\mathrm{n.-pert.}}(M^{2}),
\label{eq:DecayNP}
\end{eqnarray}%
where $\mathcal{F}^{\mathrm{n.-pert.}}(M^{2})$ is given by the formula
\begin{eqnarray}
&&\mathcal{F}^{\mathrm{n.-pert.}}(M^{2})=\Big \langle\frac{\alpha _{s}G^{2}}{%
\pi }\Big \rangle\int_{0}^{1}f_{1}(x,M^{2})dx-\Big \langle g_{s}^{3}G^{3}%
\Big \rangle  \notag \\
&&\times \int_{0}^{1}f_{2}(x,M^{2})dx-\Big \langle\frac{\alpha _{s}G^{2}}{%
\pi }\Big \rangle^{2}\int_{0}^{1}f_{3}(x,M^{2})dx.  \label{eq:NPert}
\end{eqnarray}%
The nonperturbative contributions of four, six and eight dimensions are
proportional to $\langle \alpha _{s}G^{2}/\pi \rangle $, $\langle
g_{s}^{3}G^{3}\rangle $ and $\langle \alpha _{s}G^{2}/\pi \rangle ^{2}$,
respectively. The functions $f_{i}(x,M^{2})$, $i=1,2,3$ are explicitly given
below:
\begin{eqnarray}
&&f_{1}(x,M^{2})=\frac{1}{18M^{4}x^{2}(1-x)^{2}}\left[ 8m_{c}^{2}(1-x)^{2}%
\right.  \notag \\
&&\left. +M^{2}(2-7x+9x^{2}-4x^{3}+2x^{4})\right] e^{-m_{c}^{2}/M^{2}x(1-x)},
\notag \\
&&  \label{eq:NPert1}
\end{eqnarray}%
\begin{eqnarray}
&&f_{2}(x,M^{2})=\frac{1}{240M^{8}\pi ^{2}x^{5}(x-1)^{5}}\left[
2M^{4}x^{2}(1-x)^{2}\right.  \notag \\
&&\times \left( 3-11x+15x^{2}-8x^{3}+4x^{4}\right) +24m_{c}^{4}(1-2x)^{2}
\notag \\
&&\times \left( -2-7x+17x^{2}-20x^{3}+10x^{4}\right) -3m_{c}^{2}M^{2}x
\notag \\
&&\times \left( 4-49x+176x^{2}-293x^{3}+218x^{4}-26x^{5}\right.  \notag \\
&&\left. \left. -40x^{6}+10x^{7}\right) \right] e^{-m_{c}^{2}/M^{2}x(1-x)},
\label{eq:NPert2}
\end{eqnarray}%
and%
\begin{eqnarray}
f_{3}(x,M^{2}) &=&\frac{16\pi ^{2}m_{c}^{2}}{9M^{10}x^{3}(x-1)^{3}}\left(
25m_{c}^{2}+6M^{2}x\right.  \notag \\
&&\left. -6M^{2}x^{2}\right) e^{-m_{c}^{2}/M^{2}x(1-x)}.  \label{eq:NPert3}
\end{eqnarray}%
The width of the process $X\rightarrow J/\psi \phi $ is determined by the
formula%
\begin{equation}
\Gamma (X\rightarrow J/\psi \phi )=G^{2}\frac{\lambda (m,m_{1},m_{\phi })}{%
24\pi m^{2}}|M|^{2},
\end{equation}%
where
\begin{eqnarray}
&&|M|^{2}=\frac{1}{4m^{2}m_{1}^{2}m_{\phi }^{2}}\left[
m_{1}^{8}+8m_{1}^{6}(m^{2}+m_{\phi }^{2})\right.  \notag \\
&&+(m_{\phi }^{2}-m^{2})^{2}\left( m_{\phi }^{4}+10m_{\phi
}^{2}m^{2}+m^{4}\right)  \notag \\
&&-2m_{1}^{4}\left( 9m_{\phi }^{4}+16m_{\phi }^{2}m^{2}+9m^{4}\right)
+8m_{1}^{2}  \notag \\
&&\left. \times \left( m_{\phi }^{6}-4m_{\phi }^{4}m^{2}-4m_{\phi
}^{2}m^{4}+m^{6}\right) \right] ,
\end{eqnarray}%
and $\lambda (a,b,c)$ is the function%
\begin{equation}
\lambda (a,b,c)=\frac{\sqrt{%
a^{4}+b^{4}+c^{4}-2(a^{2}b^{2}+a^{2}c^{2}+b^{2}c^{2})}}{2a}.
\end{equation}

\begin{table}[tbp]
\begin{tabular}{|c|c|}
\hline\hline
Parameters & Values (in $\mathrm{MeV}$ units) \\ \hline\hline
$m_1[m_{J/\psi}]$ & $3096.900 \pm 0.006$ \\
$f_1[f_{J/\psi}]$ & $409 \pm 15$ \\
$m_2[m_{\eta_c}]$ & $2983.9 \pm 0.5$ \\
$f_2[f_{\eta_c}]$ & $320 \pm 40$ \\
$m_3[m_{\chi_{1c}}]$ & $3510.67 \pm 0.05$ \\
$f_3[f_{\chi_{1c}}]$ & $344 \pm 27$ \\
$m_{\phi}$ & $1019.461 \pm 0.019$ \\
$f_{\phi}$ & $228.5 \pm 3.6 $ \\
$m_{\eta}$ & $547.862 \pm 0.017$ \\
$m_{\eta^{\prime}} $ & $957.78 \pm 0.06 $ \\ \hline\hline
\end{tabular}%
\caption{Masses and decay constants of mesons, which have been used in
numerical computations. }
\label{tab:Param}
\end{table}

The sum rule Eq.\ (\ref{eq:SR1}) depends on the mass and decay constant of
the $J/\psi $ and $\phi $ mesons: Their values are collected in Table\ \ref%
{tab:Param}. This table contains also spectroscopic parameters of other
mesons which will be used in the next section. The masses of all mesons are
borrowed from Ref.\ \cite{PDG:2020}. As the decay constants $f_{\phi }$ and $%
f_{1}$ of the vector mesons $\phi $ and $J/\psi $, we use their experimental
values reported in Refs. \cite{Chakraborty:2017hry,Kiselev:2001xa},
respectively. For the decay constants $f_{2}$ and $f_{3}$ of the $\eta _{c}$
and $\chi _{1c}$ mesons, we utilize relevant sum rules' predictions from
Refs.\ \cite{Colangelo:1992cx} and \cite{VeliVeliev:2012cc}, respectively.

\begin{figure}[h]
\includegraphics[width=8.8cm]{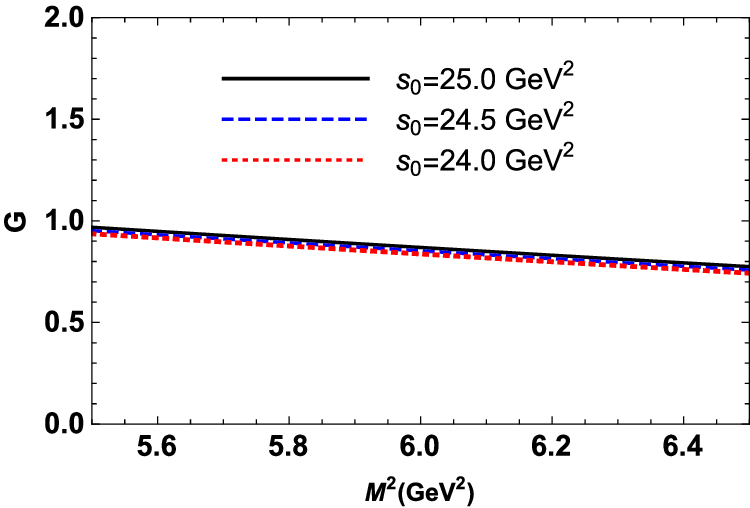}
\caption{The strong coupling $G$ as a function of the Borel parameter $M^{2}$
at fixed $s_{0}$.}
\label{fig:coupling}
\end{figure}

In numerical analysis, the parameters $M^{2}$ and $s_{0}$ are chosen as in
Eq.\ (\ref{eq:Regions}). Computations allow us to find numerical value of
the strong coupling $G$
\begin{equation}
G=0.85\pm 0.12.  \label{eq:Coupling0}
\end{equation}%
In Fig.\ \ref{fig:coupling}, we depict $G$ as a function of the Borel
parameter $M^{2}$ at fixed $s_{0}$. One sees that, the coupling $G$ is
sensitive to $M^{2}$ and $s_{0}$, which are main sources of theoretical
ambiguities of the analysis: Ambiguities arising due to variations of the
parameters $M^{2}$ and $s_{0}$ are equal to $\Delta ^{\mathrm{(M}^{\mathrm{2}%
}\mathrm{,s}_{\mathrm{0}}\mathrm{)}}G=\pm 0.11$. Uncertainties in the decay
constants $f_{1}$ and $f_{\phi }$ generates $\Delta ^{\mathrm{(f}_{\mathrm{1}%
}\mathrm{)}}G=\pm 0.03$ and $\Delta ^{\mathrm{(f}_{\mathrm{\phi }}\mathrm{)}%
}G=\pm 0.02$, respectively. Errors connected with various vacuum condensates
are very small and can be neglected.

For the partial width of the process $X\rightarrow J/\psi \phi $, we get
\begin{equation}
\Gamma (X\rightarrow J/\psi \phi )=(113\pm 30)~\mathrm{MeV}.  \label{eq:DW1}
\end{equation}%
This information will be used below to evaluate full width of the tetraquark
$X$.

\section{Processes $X\to \protect\eta _{c} \protect\eta ^{(\prime )}$ and $%
X\to \protect\chi _{1c}\protect\eta ^{(\prime )}$}

\label{sec:Decay2}
In this section, we consider processes $X\rightarrow \eta _{c}\eta ^{(\prime
)}$ and $X\rightarrow \chi _{1c}\eta ^{(\prime )}$ and calculate their
partial widths. It is not difficult to see, that decays to pseudoscalar
mesons $\eta _{c}(1S)$ and $\eta ^{(\prime )}$ with spin-parities $J^{%
\mathrm{PC}}=0^{-+}$ are $P$-wave modes of the tetraquark $X$. The second
pair of decays to axial-vector meson $\chi _{1c}(1P)$ with $J^{\mathrm{PC}%
}=1^{++}$ and $\eta ^{(\prime )}$ are its $S$-wave modes. In all of these
processes conservation of $C$-parity is the case.

\subsection{Decays $X\rightarrow \protect\eta _{c}\protect\eta $ and $%
X\rightarrow \protect\eta _{c}\protect\eta ^{\prime }$}

We start from analysis of the processes $X\rightarrow \eta _{c}\eta $ and $%
X\rightarrow \eta _{c}\eta ^{\prime }$, and extract couplings $g_{1}$ and $%
g_{2}$ which describe strong interaction at the vertices $X\eta _{c}\eta $
and $X\eta _{c}\eta ^{\prime }$, respectively.

The strong coupling $g_{1}$ is defined through on-mass-shell matrix element
\begin{equation}
\langle \eta (q)\eta _{c}\left( p\right) |X(p^{\prime })\rangle =g_{1}p\cdot
\varepsilon (p^{\prime }).  \label{eq:G2}
\end{equation}%
The correlation function for analysis of this coupling has the following
form
\begin{equation}
\widetilde{\Pi }_{\mu }(p,q)=i\int d^{4}xe^{ipx}\langle \eta (q)|\mathcal{T}%
\{J^{\eta _{c}}(x)J_{\mu }^{\dag }(0)\}|0\rangle ,  \label{eq:CorrF6}
\end{equation}%
where $J^{\eta _{c}}(x)$ is the interpolating current for the $\eta _{c}$
meson%
\begin{equation}
J^{\eta _{c}}(x)=\overline{c}_{l}(x)i\gamma _{5}c_{l}(x).
\end{equation}%
The term which will be used to determine $g_{1}$ is
\begin{eqnarray}
&&\widetilde{\Pi }_{\mu }^{\mathrm{Phys}}(p,q)=g_{1}\frac{fm_{2}^{2}f_{2}}{%
4m_{c}m\left( p^{\prime 2}-m^{2}\right) \left( p^{2}-m_{2}^{2}\right) }
\notag \\
&&\times \left[ (m_{2}^{2}-m_{\eta }^{2}-m^{2})p_{\mu }+(m_{2}^{2}+m_{\eta
}^{2}+m^{2})q_{\mu }\right]  \notag \\
&&+\cdots .  \label{eq:CorrF8}
\end{eqnarray}%
Here, $m_{2}$ and $m_{\eta }$ are masses of the $\eta _{c}$ and $\eta $
mesons, respectively. The decay constant of the $\eta _{c}$ meson is denoted
by $f_{2}$. Let us note that to derive Eq.\ (\ref{eq:CorrF8}), we use the
matrix elements of the tetraquark $X$ from Eq.\ (\ref{eq:Res}), and the
matrix element of the $\eta _{c}$ meson
\begin{equation}
\langle 0|J^{\eta _{c}}|\eta _{c}\left( p\right) \rangle =\frac{%
f_{2}m_{2}^{2}}{2m_{c}}.  \label{eq:MatEl}
\end{equation}

The QCD side of the sum rule reads%
\begin{eqnarray}
&&\widetilde{\Pi }_{\mu }^{\mathrm{OPE}}(p,q)=i\int d^{4}xe^{ipx}\epsilon
\widetilde{\epsilon }\left\{ \left[ \gamma _{5}\widetilde{S}%
_{c}^{lc}(x)\gamma _{5}\right. \right.  \notag \\
&&\left. \left. \times \widetilde{S}_{c}^{el}(-x)\gamma _{\mu }\gamma _{5}%
\right] +\left[ \gamma _{\mu }\gamma _{5}\widetilde{S}_{c}^{lc}(x){}\gamma
_{5}\widetilde{S}_{c}^{el}(-x){}\gamma _{5}\right] \right\} _{\alpha \beta }
\notag \\
&&\times \langle \eta (q)|\overline{s}_{\alpha }^{b}(0)s_{\beta
}^{d}(0)|0\rangle .  \label{eq:CorrF9}
\end{eqnarray}%
It is clear that $\widetilde{\Pi }_{\mu }^{\mathrm{OPE}}(p,q)$ contains only
local matrix elements of $\eta $, therefore remaining calculations have to
be carried out in the context of the soft-meson approximation. Technical
methods of such treatment have been explained in the previous section.
Therefore, we do not concentrate on further details, and note that in the
soft limit $\widetilde{\Pi }_{\mu }^{\mathrm{OPE}}(p,0)$ receives
contributions only from the matrix element
\begin{equation}
2m_{s}\langle \eta |\overline{s}i\gamma _{5}s|0\rangle =h_{\eta }^{s}.
\label{eq:H}
\end{equation}%
The parameter $h_{\eta }^{s}$ in Eq.\ (\ref{eq:H}) can be defined
theoretically \cite{Agaev:2014wna}, but for our purposes it is enough to use
its phenomenological value extracted from analysis of relevant exclusive
processes. Thus, we have
\begin{equation}
h_{\eta }^{s}=-h_{s}\sin \varphi ,\ h_{s}=(0.087\pm 0.006)~\mathrm{GeV}^{3},
\end{equation}%
where $\varphi =39.3^{\circ }\pm 1.0^{\circ }$ is the $\eta -\eta ^{\prime }$
mixing angle in the quark-flavor basis (for details, see Ref.\ \cite%
{Agaev:2014wna}).

The sum rule for $g_{1}$ is derived by making use of invariant amplitudes
corresponding to structures $p_{\mu }$ in $\widetilde{\Pi }_{\mu }^{\mathrm{%
Phys}}(p)$ and $\widetilde{\Pi }_{\mu }^{\mathrm{OPE}}(p)$. It reads%
\begin{equation}
g_{1}=\frac{4mm_{c}}{fm_{2}^{2}f_{2}}\frac{\mathcal{P}(M^{2},m^{\prime 2})%
\mathcal{B}\widetilde{\Pi }^{\mathrm{OPE}}(p^{2})}{m_{2}^{2}-m^{2}-m_{\eta
}^{2}},
\end{equation}%
where $m^{\prime 2}=(m^{2}+m_{2}^{2})/2$. The Borel transformed and
subtracted invariant amplitude $\mathcal{B}\widetilde{\Pi }^{\mathrm{OPE}%
}(p^{2})$ is given by the following expression
\begin{eqnarray}
\mathcal{B}\widetilde{\Pi }^{\mathrm{OPE}}(p^{2}) &=&-\frac{h_{\eta
}^{s}m_{c}}{4\pi ^{2}m_{s}}\int_{4\mathcal{M}^{2}}^{s_{0}}ds\frac{\sqrt{%
s(s-4m_{c}^{2})}}{s}e^{-s/M^{2}}  \notag \\
&&+\widetilde{\Pi }(M^{2}).
\end{eqnarray}%
The nonperturbative component $\widetilde{\Pi }(M^{2})$ is calculated with
dimension-$8$ accuracy and determined by formulas similar to ones from Eqs.\
(\ref{eq:NPert})-(\ref{eq:NPert3}). Therefore, there is no need to write
down their explicit expressions.

The width of the mode $X\rightarrow \eta _{c}\eta $ can be calculated using
the formula
\begin{equation}
\Gamma (X\rightarrow \eta _{c}\eta )=g_{1}^{2}\frac{\lambda
^{3}(m,m_{2},m_{\eta })}{24\pi m^{2}}.  \label{eq:DW2}
\end{equation}%
Numerical computations yield%
\begin{equation}
|g_{1}|=3.75\pm 0.78,
\end{equation}%
and%
\begin{equation}
\Gamma (X\rightarrow \eta _{c}\eta )=(18.0\pm 5.4)~\mathrm{MeV.}
\end{equation}

The second process $X\rightarrow \eta _{c}\eta ^{\prime }$ can be considered
in a similar manner, difference being in the matrix element of the $\eta
^{\prime }$ meson
\begin{equation}
2m_{s}\langle \eta ^{\prime }|\overline{s}i\gamma _{5}s|0\rangle =h_{s}\cos
\varphi ,  \label{eq:MEl1}
\end{equation}%
that contributes to the corresponding correlation function. For this decay,
we find
\begin{equation}
g_{2}=4.38\pm 0.90,
\end{equation}%
and%
\begin{equation}
\Gamma (X\rightarrow \eta _{c}\eta ^{\prime })=(15.5\pm 4.5)~\mathrm{MeV}.
\label{eq:DW3}
\end{equation}%
\newline
Effects of these processes on the full width of $X$ are not small, and will
be taken into account.

\subsection{Decays $X\rightarrow \protect\chi _{1c}\protect\eta $ and $%
X\rightarrow \protect\chi _{1c}\protect\eta ^{\prime }$}

Processes $X\rightarrow \chi _{1c}\eta $ and $X\rightarrow \chi _{1c}\eta
^{\prime }$ are explored in accordance with the scheme described above.
Here, we have to evaluate the strong couplings $g_{3}$ and $g_{4}$ which
correspond to vertices $X\chi _{1c}\eta $ and $X\chi _{1c}\eta ^{\prime }$.

Let us consider the decay $X\rightarrow \chi _{1c}\eta $ and write down some
principal expressions. The relevant strong coupling $g_{3}$ is defined by
the on-mass-shell matrix element
\begin{eqnarray}
&&\langle \eta (q)\chi _{1c}\left( p\right) |X(p^{\prime })\rangle
=g_{3}\left\{ \left[ p\cdot p^{\prime }\right] \left[ \varepsilon ^{\ast
}(p)\cdot \varepsilon (p^{\prime })\right] \right.  \notag \\
&&\left. -[p\cdot \varepsilon (p^{\prime })][p^{\prime }\cdot \varepsilon
^{\ast }(p)]\right\} ,  \label{eq:Vertex}
\end{eqnarray}%
with $\varepsilon _{\mu }^{\ast }(p)$ being the polarization vector of the
meson $\chi _{1c}$.

To determine $g_{3}$, we consider the correlation function
\begin{equation}
\widehat{\Pi }_{\mu \nu }(p,q)=i\int d^{4}xe^{ipx}\langle \eta (q)|\mathcal{T%
}\{J_{\nu }^{\chi _{1c}}(x)J_{\mu }^{\dag }(0)\}|0\rangle ,
\label{eq:CorrF10}
\end{equation}%
where $J_{\mu }^{\chi _{1c}}(x)$ is the interpolating current for the
axial-vector meson $\chi _{1c}$%
\begin{equation}
J_{\nu }^{\chi _{1c}}(x)=\overline{c}_{l}(x)i\gamma _{5}\gamma _{\nu
}c_{l}(x).
\end{equation}%
Then, the term in $\widehat{\Pi }_{\mu \nu }(p,q)$ which has two poles in
variables $p^{2}$ and $p^{\prime 2}=(p+q)^{2}$ is given by the formula%
\begin{eqnarray}
&&\widehat{\Pi }_{\mu \nu }^{\mathrm{Phys}}(p,q)=g_{3}\frac{fmf_{3}m_{3}}{%
\left( p^{\prime 2}-m^{2}\right) \left( p^{2}-m_{3}^{2}\right) }  \notag \\
&&\times \left[ \frac{1}{2}\left( m^{2}+m_{3}^{2}-m_{\eta }^{2}\right)
g_{\mu \nu }-p_{\mu }p_{\nu }^{\prime }\right]  \notag \\
&&+\cdots ,  \label{eq:CorrF12a}
\end{eqnarray}%
where $m_{3}$ and $f_{3}$ are the mass and decay constant of the $\chi _{1c}$
meson. As usual, dots stand for contributions of higher resonances and
continuum states. The ground-state term in $\widehat{\Pi }_{\mu \nu }^{%
\mathrm{Phys}}(p,q)$ has been found using the matrix element

\begin{equation}
\langle 0|J_{\mu }^{\chi _{1c}}|\chi _{1c}\left( p\right) \rangle
=f_{3}m_{3}\varepsilon _{\mu }(p),  \label{eq:MEl}
\end{equation}%
as well as matrix element of the tetraquark $X$.

The correlation function $\widehat{\Pi }_{\mu \nu }^{\mathrm{OPE}}(p,q)$ is
given by the expression%
\begin{eqnarray}
&&\widehat{\Pi }_{\mu \nu }^{\mathrm{OPE}}(p,q)=i\int d^{4}xe^{ipx}\epsilon
\widetilde{\epsilon }\left\{ \left[ \gamma _{5}\widetilde{S}%
_{c}^{lc}(x)\gamma _{\nu }\gamma _{5}\right. \right.  \notag \\
&&\left. \left. \times \widetilde{S}_{c}^{el}(-x)\gamma _{\mu }\gamma _{5}%
\right] +\left[ \gamma _{\mu }\gamma _{5}\widetilde{S}_{c}^{lc}(x){}\gamma
_{\nu }\gamma _{5}\widetilde{S}_{c}^{el}(-x){}\gamma _{5}\right] \right\}
_{\alpha \beta }  \notag \\
&&\times \langle \eta (q)|\overline{s}_{\alpha }^{b}(0)s_{\beta
}^{d}(0)|0\rangle.  \label{eq:CorrF13}
\end{eqnarray}

The required sum rule for the coupling $g_{3}$ is derived by equating
invariant amplitudes of structures $\sim g_{\mu \nu }$ from the functions $%
\widehat{\Pi }_{\mu \nu }^{\mathrm{Phys}}(p,q)$ and $\widehat{\Pi }_{\mu \nu
}^{\mathrm{OPE}}(p,q)$, and has the form
\begin{equation}
g_{3}=\frac{2}{fmf_{3}m_{3}}\frac{\mathcal{P}(M^{2},\widehat{m}^{2})\mathcal{%
B}\widehat{\Pi }^{\mathrm{OPE}}(p^{2})}{2\widehat{m}^{2}-m_{\eta }^{2}},
\end{equation}%
where $\widehat{m}^{2}=(m^{2}+m_{3}^{2})/2$. Numerical analysis for $g_{3}$
gives
\begin{equation}
g_{3}=(1.34\pm 0.23)\times 10^{-1}~\mathrm{GeV}^{-1}.
\end{equation}

Width of the process $X\rightarrow \chi _{1c}\eta $ is determined by the
expression%
\begin{equation}
\Gamma (X\rightarrow \chi _{1c}\eta )=g_{3}^{2}\frac{\lambda m_{3}^{2}}{%
24\pi }\left( 3+\frac{2\lambda ^{2}}{m_{3}^{2}}\right),  \label{eq:DW4}
\end{equation}%
with $\lambda $ being equal to $\lambda (m,m_{3},m_{\eta })$. Then it is not
difficult to find that

\begin{equation}
\Gamma (X\rightarrow \chi _{1c}\eta )=(7.9\pm 1.9)~\mathrm{MeV}.
\end{equation}%
For the decay $X\rightarrow \chi _{1c}\eta ^{\prime }$, we get
\begin{equation}
|g_{4}|=(1.39\pm 0.23)\times 10^{-1}~\mathrm{GeV}^{-1},
\end{equation}%
and

\begin{equation}
\Gamma (X\rightarrow \chi _{1c}\eta ^{\prime })=(4.9\pm 1.2)\ \ \mathrm{MeV},
\end{equation}%
respectively.


\section{Summing up}

\label{sec:Conclusions}
The full width of the tetraquark $X$ can be evaluated using results for the
partial width of its five decay modes obtained in Sections \ref{sec:Decay1}
and \ref{sec:Decay2}. One of these modes $X\rightarrow J/\psi \phi $ is the
dominant decay channel of the tetraquark $X$, whereas remaining processes
are sub-dominant ones. After simple computations, we get%
\begin{equation}
\Gamma =(159\pm 31)~\mathrm{MeV}.  \label{eq:Fwd}
\end{equation}%
Our result for the full width $\Gamma $ of the tetraquark $X$ is in a very
nice agreement with $\Gamma _{\exp }=(174\pm 27_{-73}^{+134})~\mathrm{MeV}$
found by the LHCb collaboration.

But by drawing such conclusions, we take into account that both theoretical
and experimental information on the full with of $X(4630)$ suffers from
errors. The uncertainties are large in the case of \ $\Gamma _{\exp }$ which
limit credibility of conclusions that are based on these data. Experimental
errors  also make it  difficult to obtain detailed comparisons and choices between existing
theoretical models for $X(4630)$. In this sense, more precise measurements
of $\Gamma _{\exp }$ are required.

From another side, our present result can be further refined by including
into analysis other decay modes of $X$. There are a few processes which
contribute to the full width of the tetraquark $X$. Thus, decays to meson
pairs $D_{s}^{\ast \pm }D_{s0}(2317)^{\mp }$ and $D_{s}^{\pm
}D_{s1}(2460)^{\mp }$ are among kinematically allowed channels of $X$ .
These processes belong to $\mathrm{V\rightarrow V+S}$ and $\mathrm{%
V\rightarrow PS+AV}$ type $S$-wave decay modes of $X$, respectively. Their
partial widths are determined by the expression Eq.\ (\ref{eq:DW4}) with
relevant strong coupling. To make crude estimates for partial widths of
these decays, we may assume that strong couplings at corresponding
tetraquark-meson-meson vertices are a same order of $g_{3}$ ($|g_{4}|$).
Then widths of these modes are suppressed relative to decays $X\rightarrow
\chi _{1c}\eta ^{(\prime )}$, because the factor $\Lambda =m_{\ast }^{2}$ $%
\lambda \left( 3+2\lambda ^{2}/m_{\ast }^{2}\right) /24\pi $ ($m_{\ast }$ is
a mass of a heaviest final meson) is smaller for two final-state mesons of
approximately equal mass than in the case of light and heavy mesons. For
instance, in the decay $X\rightarrow \chi _{1c}\eta $ it is equal to $%
\Lambda \approx 0.44$, while we find $\Lambda \approx 0.17$ for the process $%
X\rightarrow D_{s}^{-}D_{s1}(2460)^{+}$. But these decay channels, in total,
may compensate a $15~\mathrm{MeV}$ gap between $\Gamma _{\exp }$ and $\Gamma
$.

Another interesting field of future studies is an exploration of $X(4630)$
in the molecule picture using the QCD sum rule method. This is necessary to
compare predictions for the molecule and diquark-antidiquark models with
each another, as well as with the LHCb data. In the context of the QCD sum
rule approach diquark-antidiquark and molecule models for the same resonance
lead to different results \cite{Agaev:2021vur,Agaev:2022ast}. As a rule, a
molecule of conventional mesons is heavier than a diquark-antidiquark
structure with identical content and spin-parities. A width of such molecule
is also larger than that of its diquark counterpart, i.e., a diquark
structure is more stable than a meson molecule. Nevertheless, despite
existing investigations of $X(4630)$ in different approaches, it is
necessary to examine the molecule model for $X(4630)$ in the context of the
QCD sum rule method as well.

Analysis performed in the present article and gained knowledge about the
mass and full width of the tetraquark $X$, as well as a very nice agreement
between these parameters and LHCb measurements allows us to interpret $%
X(4630)$ as the vector diquark-antidiquark state $X$ with the spin-parities $%
J^{\mathrm{PC}}=1^{-+}$.

\section*{ACKNOWLEDGMENTS}

S.~S.~A. is grateful to Prof. V.~M.~Braun for enlightening comments on some
items in the LCSR method.

\begin{widetext}

\appendix*

\section{ The quark propagators and invariant amplitude $\Pi (M^{2},s_{0})$}

\renewcommand{\theequation}{\Alph{section}.\arabic{equation}} \label{sec:App}

In the current article, for the light quark propagator $S_{q}^{ab}(x)$, we
employ the following expression
\begin{eqnarray}
&&S_{q}^{ab}(x)=i\delta _{ab}\frac{\slashed x}{2\pi ^{2}x^{4}}-\delta _{ab}%
\frac{m_{q}}{4\pi ^{2}x^{2}}-\delta _{ab}\frac{\langle \overline{q}q\rangle
}{12}+i\delta _{ab}\frac{\slashed xm_{q}\langle \overline{q}q\rangle }{48}%
-\delta _{ab}\frac{x^{2}}{192}\langle \overline{q}g_{s}\sigma Gq\rangle
\notag \\
&&+i\delta _{ab}\frac{x^{2}\slashed xm_{q}}{1152}\langle \overline{q}%
g_{s}\sigma Gq\rangle -i\frac{g_{s}G_{ab}^{\alpha \beta }}{32\pi ^{2}x^{2}}%
\left[ \slashed x{\sigma _{\alpha \beta }+\sigma _{\alpha \beta }}\slashed x%
\right] -i\delta _{ab}\frac{x^{2}\slashed xg_{s}^{2}\langle \overline{q}%
q\rangle ^{2}}{7776}  \notag \\
&&-\delta _{ab}\frac{x^{4}\langle \overline{q}q\rangle \langle
g_{s}^{2}G^{2}\rangle }{27648}+\cdots .
\end{eqnarray}%
For the heavy quark $Q=c$, we use the propagator $S_{Q}^{ab}(x)$
\begin{eqnarray}
&&S_{Q}^{ab}(x)=i\int \frac{d^{4}k}{(2\pi )^{4}}e^{-ikx}\Bigg \{\frac{\delta
_{ab}\left( {\slashed k}+m_{Q}\right) }{k^{2}-m_{Q}^{2}}-\frac{%
g_{s}G_{ab}^{\alpha \beta }}{4}\frac{\sigma _{\alpha \beta }\left( {\slashed %
k}+m_{Q}\right) +\left( {\slashed k}+m_{Q}\right) \sigma _{\alpha \beta }}{%
(k^{2}-m_{Q}^{2})^{2}}  \notag \\
&&+\frac{g_{s}^{2}G^{2}}{12}\delta _{ab}m_{Q}\frac{k^{2}+m_{Q}{\slashed k}}{%
(k^{2}-m_{Q}^{2})^{4}}+\frac{g_{s}^{3}G^{3}}{48}\delta _{ab}\frac{\left( {%
\slashed k}+m_{Q}\right) }{(k^{2}-m_{Q}^{2})^{6}}\left[ {\slashed k}\left(
k^{2}-3m_{Q}^{2}\right) +2m_{Q}\left( 2k^{2}-m_{Q}^{2}\right) \right] \left(
{\slashed k}+m_{Q}\right) +\cdots \Bigg \}.  \notag \\
&&
\end{eqnarray}

Here, we have used the short-hand notations
\begin{equation}
G_{ab}^{\alpha \beta }\equiv G_{A}^{\alpha \beta }\lambda _{ab}^{A}/2,\ \
G^{2}=G_{\alpha \beta }^{A}G_{A}^{\alpha \beta },\ G^{3}=f^{ABC}G_{\alpha
\beta }^{A}G^{B\beta \delta }G_{\delta }^{C\alpha },
\end{equation}%
where $G_{A}^{\alpha \beta }$ is the gluon field strength tensor, $\lambda
^{A}$ and $f^{ABC}$ are the Gell-Mann matrices and structure constants of
the color group $SU_{c}(3)$, respectively. The indices $A,B,C$ run in the
range $1,2,\ldots 8$.

The invariant amplitude $\Pi (M^{2},s_{0})$ obtained after the Borel
transformation and subtraction procedures is given by the expression%
\begin{equation*}
\Pi (M^{2},s_{0})=\int_{4\mathcal{M}^{2}}^{s_{0}}ds\rho ^{\mathrm{OPE}%
}(s)e^{-s/M^{2}}+\Pi (M^{2}),
\end{equation*}%
where the spectral density $\rho ^{\mathrm{OPE}}(s)$ and the function $\Pi
(M^{2})$ are determined by formulas
\begin{equation}
\rho ^{\mathrm{OPE}}(s)=\rho ^{\mathrm{pert.}}(s)+\sum_{N=3}^{8}\rho ^{%
\mathrm{DimN}}(s),\ \ \Pi (M^{2})=\sum_{N=6}^{10}\Pi ^{\mathrm{DimN}}(M^{2}),
\label{eq:A1}
\end{equation}%
respectively. The components of $\rho ^{\mathrm{OPE}}(s)$ and $\Pi (M^{2})$
are given by the expressions%
\begin{equation}
\rho ^{\mathrm{DimN}}(s)=\int_{0}^{1}d\alpha \int_{0}^{1-a}d\beta \rho ^{%
\mathrm{DimN}}(s,\alpha ,\beta ),\ \ \Pi ^{\mathrm{DimN}}(M^{2})=%
\int_{0}^{1}d\alpha \int_{0}^{1-a}d\beta \Pi ^{\mathrm{DimN}}(M^{2},\alpha
,\beta ) , \label{eq:A2}
\end{equation}%
if $\rho ^{\mathrm{DimN}}(s,\alpha ,\beta )$ and $\Pi ^{\mathrm{DimN}%
}(M^{2},\alpha ,\beta )$ are functions of $\alpha $ and $\beta $, and by
formulas
\begin{equation}
\rho ^{\mathrm{DimN}}(s)=\int_{0}^{1}d\alpha \rho ^{\mathrm{DimN}}(s,\alpha
),\ \ \Pi ^{\mathrm{DimN}}(M^{2})=\int_{0}^{1}d\alpha \Pi ^{\mathrm{DimN}%
}(M^{2},\alpha ),  \label{eq:A4}
\end{equation}%
provided that  they depend only on $\alpha $. Let us note that in Eqs.\ (\ref%
{eq:A2}) and (\ref{eq:A4}) variables $\alpha $ and $\beta $ are Feynman
parameters.

The perturbative and nonperturbative components of the spectral density $%
\rho ^{\mathrm{pert.}}(s,\alpha ,\beta )$ and $\rho ^{\mathrm{Dim3(4,5,6,7,8)%
}}(s,\alpha ,\beta )$ \ have the forms:
\begin{eqnarray}
&&\rho ^{\mathrm{pert.}}(s,\alpha ,\beta )=\frac{\Theta (L_{1})}{1536\pi
^{6}L^{2}N_{1}^{8}}\left[ m_{c}^{2}N_{2}-s\alpha \beta L\right] ^{2}\left\{
12m_{c}^{3}m_{s}L(\alpha +\beta )^{2}N_{1}^{3}-m_{c}^{4}\alpha \beta
N_{1}^{2}\left[ 5\beta ^{3}+5\alpha ^{2}(\alpha -1)\right. \right.  \notag \\
&&\left. +\alpha \beta (-10+13\alpha )+\beta ^{2}(-5+13\alpha )\right]
-35s^{2}\alpha ^{3}\beta ^{3}L^{3}-12m_{c}m_{s}s\alpha \beta (\alpha +\beta )%
\left[ \beta ^{3}+2\beta ^{2}(\alpha -1)+\alpha (\alpha -1)^{2}\right.
\notag \\
&&\left. +\beta (1-3\alpha +2\alpha ^{2})\right] ^{2}+2m_{c}^{2}s\alpha
^{2}\beta ^{2}\left[ \beta ^{3}+2\beta ^{2}(\alpha -1)+\alpha (\alpha
-1)^{2}+\beta (1-3\alpha +2\alpha ^{2})\right] \left[ 14\beta ^{2}+14\alpha
(\alpha -1)\right.  \notag \\
&&\left. \left. +\beta (-14+27\alpha )\right] \right\} ,
\end{eqnarray}%
\begin{eqnarray}
&&\rho ^{\mathrm{Dim3}}(s,\alpha ,\beta )=-\frac{\langle \overline{s}%
s\rangle \Theta (L_{1})}{16\pi ^{4}N_{1}^{6}}\left\{
m_{c}^{5}N_{2}^{3}-m_{s}s^{2}\alpha ^{2}\beta ^{2}L^{3}\left[ \beta
^{2}+\alpha (\alpha -1)-\beta (1+14\alpha )\right] +m_{c}s^{2}\alpha
^{2}\beta ^{2}L^{2}N_{2}\right.  \notag \\
&&-2m_{c}^{3}s\alpha \beta LN_{2}^{2}+3m_{c}^{4}m_{s}N_{1}^{2}\left[ \beta
^{5}+\alpha ^{3}(\alpha -1)^{2}+\beta ^{4}(-2+4\alpha )+\beta \alpha
^{2}(3-7\alpha +4\alpha ^{2})+\beta ^{2}\alpha (3-10\alpha +8\alpha
^{2})\right.  \notag \\
&&\left. +\beta ^{3}(1-7\alpha +8\alpha ^{2})\right] -sm_{c}^{2}m_{s}\alpha
\beta \left[ 2\beta ^{7}+2\alpha ^{3}(\alpha -1)^{4}-\beta ^{6}(8-23\alpha
)+\beta \alpha ^{2}(\alpha -1)^{3}(-6+23\alpha )\right.  \notag \\
&&+\beta ^{5}(12-75\alpha +79\alpha ^{2})+\beta ^{2}\alpha (\alpha
-1)^{2}(6-54\alpha +79\alpha ^{2})+\beta ^{4}(-8+87\alpha -212\alpha
^{2}+133\alpha ^{3})  \notag \\
&&\left. \left. +\beta ^{3}(2-41\alpha +193\alpha ^{2}-287\alpha
^{3}+133\alpha ^{4})\right] \right\} ,
\end{eqnarray}%
\begin{eqnarray}
&&\rho ^{\mathrm{Dim4}}(s,\alpha ,\beta )=-\frac{\langle \alpha
_{s}G^{2}/\pi \rangle \Theta (L_{1})}{9216\pi ^{4}L^{2}N_{1}^{6}}\left\{
-15s^{2}\alpha ^{3}\beta ^{3}L^{4}(9\beta +4\alpha )+m_{c}^{4}\alpha \beta
N_{1}^{2}\left[ 2\beta ^{5}+\beta ^{4}(28-82\alpha )+\beta ^{2}\alpha \left(
-72\right. \right. \right.  \notag \\
&&\left. \left. +255\alpha -163\alpha ^{2}\right) +\beta \alpha
^{2}(-54+97\alpha -19\alpha ^{2})+4\alpha ^{3}(-3-2\alpha +5\alpha
^{2})-2\beta ^{3}(15-89\alpha +104\alpha ^{2})\right]  \notag \\
&&+6sm_{c}m_{s}\alpha \beta L^{2}\left[ 5\beta ^{7}+2\alpha ^{4}(\alpha
-1)^{2}(3+\alpha )-2\beta \alpha ^{3}(\alpha -1)^{2}(-28+31\alpha )-\beta
^{6}(7+53\alpha )-\beta ^{2}\alpha ^{2}(\alpha -1)^{2}\right.  \notag \\
&&\left. \times (-97+173\alpha )+\beta ^{5}(-1+156\alpha -161\alpha
^{2})+\beta ^{3}\alpha (50-355\alpha +588\alpha ^{2}-283\alpha ^{3})+\beta
^{4}(3-153\alpha +419\alpha ^{2}-277\alpha ^{3})\right]  \notag \\
&&-6m_{c}^{3}m_{s}N_{1}^{2}\left[ 5\beta ^{7}+2\alpha ^{4}(\alpha
-1)^{2}(3+\alpha )-\beta ^{6}(7+26\alpha )+\beta ^{5}(-1+96\alpha -143\alpha
^{2})+\beta ^{4}(3-106\alpha +356\alpha ^{2}-259\alpha ^{3})\right.  \notag
\\
&&\left. +\beta ^{2}\alpha ^{2}(69-297\alpha +389\alpha ^{2}-161\alpha
^{3})+2\beta ^{3}\alpha (18-141\alpha +255\alpha ^{2}-134\alpha ^{3})-2\beta
\alpha ^{3}(-21+65\alpha -63\alpha ^{2}+19\alpha ^{3})\right]  \notag \\
&&+m_{c}^{2}s\alpha ^{2}\beta ^{2}\left[ \beta ^{3}+2\beta ^{2}(\alpha
-1)+\alpha (\alpha -1)^{2}+\beta (1-3\alpha +2\alpha ^{2})\right] \left[
75\beta ^{4}+3\beta ^{3}(-74+149\alpha )-12\alpha ^{2}(-5+4\alpha +\alpha
^{2})\right.  \notag \\
&&\left. \left. +\alpha \beta (207-535\alpha +264\alpha ^{2})+\beta
^{2}(147-718\alpha +627\alpha ^{2})\right] \right\} ,
\end{eqnarray}%
\begin{eqnarray}
&&\rho _{1}^{\mathrm{Dim5}}(s,\alpha ,\beta )=\frac{\langle \overline{s}%
g_{s}\sigma Gs\rangle \Theta (L_{1})L}{96\pi ^{4}N_{1}^{5}}\left\{
-16sm_{s}\alpha ^{2}\beta ^{2}L^{2}+3m_{c}^{3}N_{2}^{2}-3sm_{c}\alpha \beta %
\left[ \beta ^{4}+\alpha ^{2}(\alpha -1)^{2}+\beta ^{3}(-2+3\alpha )\right.
\right.  \notag \\
&&\left. +\alpha \beta (2-5\alpha +3\alpha ^{2})+\beta ^{2}(1-5\alpha
+4\alpha ^{2})\right] +m_{c}^{2}m_{s}\alpha \beta \left[ 7\beta ^{4}+7\alpha
^{2}(\alpha -1)^{2}+2\beta ^{3}(-7+10\alpha )\right.  \notag \\
&&\left. \left. +2\alpha \beta (7-17\alpha +10\alpha ^{2})+\beta
^{2}(7-34\alpha +27\alpha ^{2})\right] \right\} ,
\end{eqnarray}%
\begin{eqnarray}
&&\rho _{1}^{\mathrm{Dim6}}(M^{2},\alpha ,\beta )=-\frac{\Theta (L_{1})}{%
405\cdot 2^{12}\pi ^{6}(\beta -1)^{2}L^{2}N_{1}^{7}}\left\{ 2560g_{s}^{2}\pi
^{2}\langle \overline{s}s\rangle ^{2}(\beta -1)^{2}\alpha \beta
L^{3}N_{1}^{2}\left[ -16s\alpha \beta L^{2}+m_{c}^{2}\left( 7\beta
^{4}+7\alpha ^{2}\right. \right. \right.  \notag \\
&&\left. \left. \times (\alpha -1)^{2}+2\beta ^{3}(-7+10\alpha )+2\beta
\alpha (7-17\alpha +10\alpha ^{2})+\beta ^{2}(7-34\alpha +27\alpha
^{2})\right) \right] +9\langle g_{s}^{3}G^{3}\rangle \left[
-18m_{c}m_{s}(\beta -1)^{2}N_{1}^{2}\right.  \notag \\
&&\left( 3\beta ^{9}+\beta ^{8}(-9+\alpha )-3\beta ^{3}\alpha ^{5}(\alpha
-1)-5\beta ^{2}\alpha ^{5}(\alpha -1)^{2}+3\alpha ^{6}(\alpha -1)^{3}+\beta
^{7}(9-5\alpha ^{2})+\beta ^{5}\alpha (2-5\alpha +3\alpha ^{2})\right.
\notag \\
&&\left. +\beta \alpha ^{5}(2-3\alpha +\alpha ^{3})-\beta ^{6}(3+3\alpha
-10\alpha ^{2}+3\alpha ^{3})\right) +3m_{c}^{2}\alpha \beta \left( 17\beta
^{13}+2\beta ^{12}(-51+19\alpha )-\alpha ^{7}(\alpha -1)^{4}(-17-2\alpha
+\alpha ^{2})\right.  \notag \\
&&+\beta ^{11}(255-214\alpha +37\alpha ^{2})-\beta ^{10}(340-500\alpha
+182\alpha ^{2}+\alpha ^{3})-\beta ^{2}\alpha ^{5}(\alpha
-1)^{3}(-3-79\alpha +65\alpha ^{2}+4\alpha ^{3})-\beta \alpha ^{6}(\alpha
-1)^{3}  \notag \\
&&\times (24-66\alpha +20\alpha ^{2}+7\alpha ^{3})-\beta ^{3}\alpha
^{4}(\alpha -1)^{2}(10-35\alpha -63\alpha ^{2}+71\alpha ^{3})+\beta
^{9}(255-620\alpha +355\alpha ^{2}+31\alpha ^{3}-41\alpha ^{4})  \notag \\
&&-\beta ^{4}\alpha ^{3}(\alpha -1)^{2}(10-66\alpha +107\alpha ^{2}-17\alpha
^{3}+9\alpha ^{4})-2\beta ^{8}\left( 51-215\alpha +170\alpha ^{2}+62\alpha
^{3}-105\alpha ^{4}+30\alpha ^{5}\right)  \notag \\
&&+\beta ^{7}(17-158\alpha +155\alpha ^{2}+206\alpha ^{3}-440\alpha
^{4}+282\alpha ^{4}-66\alpha ^{6})+\beta ^{6}\alpha \left( 24-22\alpha
-169\alpha ^{2}+480\alpha ^{3}-532\alpha ^{4}+281\alpha ^{5}\right.  \notag
\\
&&\left. \left. -62\alpha ^{6}\right) +\beta ^{5}\alpha ^{2}\left(
-3+67\alpha -285\alpha ^{2}+507\alpha ^{3}-449\alpha ^{4}+202\alpha
^{5}-39\alpha ^{6}\right) \right) +2s\alpha ^{2}\beta ^{2}L^{2}\left(
21\beta ^{9}-4\beta ^{8}(21+8\alpha )\right.  \notag \\
&&+\beta ^{7}(126+128\alpha -28\alpha ^{2})-3\alpha ^{5}(\alpha
-1)^{2}(-7-2\alpha +\alpha ^{2})-4\beta \alpha ^{4}(\alpha -1)^{2}(8+9\alpha
+3\alpha ^{2})  \notag \\
&&+4\beta ^{6}(-21-48\alpha +20\alpha ^{2}+3\alpha ^{3})+\beta ^{4}\alpha
(-32+16\alpha +48\alpha ^{2}-31\alpha ^{4})+2\beta ^{2}\alpha ^{3}\left(
2+44\alpha -75\alpha ^{2}+26\alpha ^{3}+3\alpha ^{4})\right.  \notag \\
&&\left. \left. \left. -4\beta ^{3}\alpha ^{2}(-1+6\alpha +17\alpha
^{2}-33\alpha ^{3}+11\alpha ^{4})+\beta ^{5}(21+128\alpha -72\alpha
^{2}-4\alpha ^{3}+12\alpha ^{4})\right) \right] \right\},
\end{eqnarray}%
\begin{eqnarray}
&&\rho _{1}^{\mathrm{Dim7}}(M^{2},\alpha ,\beta )=-\frac{\langle \alpha
_{s}G^{2}/\pi \rangle \langle \overline{s}s\rangle \Theta (L_{1})}{1152\pi
^{2}N_{1}^{4}}\left\{ 9m_{s}\beta \alpha ^{2}L^{2}+2m_{c}\left[ 5\beta
^{5}-2\beta ^{4}(1+11\alpha )+\beta ^{3}(-3+37\alpha -37\alpha ^{2})\right.
\right.  \notag \\
&& \left. \left. +\beta ^{2}\alpha (-15+48\alpha -37\alpha ^{2})+\beta
\alpha ^{2}(-15+37\alpha -22\alpha ^{2})+\alpha ^{3}(-3-2\alpha +5\alpha
^{2}) \right] \right\} ,
\end{eqnarray}%
\begin{equation}
\rho _{1}^{\mathrm{Dim8}}(M^{2},\alpha ,\beta )=\frac{5\langle \alpha
_{s}G^{2}/\pi \rangle ^{2}}{3072\pi ^{2}N_{1}^{4}}\Theta (L_{1})\alpha
^{2}\beta ^{2}L.
\end{equation}%
The spectral densities $\rho ^{\mathrm{Dim5(6,7,8)}}(s,\alpha )$ are given
by the formulas

\begin{equation}
\rho _{1}^{\mathrm{Dim5}}(s,\alpha )=-\frac{6\langle \overline{s}g_{s}\sigma
Gs\rangle m_{c}^{2}m_{s}}{96\pi ^{4}}\Theta (L_{2}),\ \
\end{equation}%
\begin{equation}
\rho _{2}^{\mathrm{Dim6}}(s,\alpha )=-\frac{\langle \overline{s}s\rangle ^{2}%
}{1296\pi ^{4}}\Theta (L_{2})\left[ 2g_{s}^{2}m_{c}m_{s}+27\pi
^{2}(8m_{c}^{2}-4m_{c}m_{s})\right] ,\ \
\end{equation}%
\begin{equation}
\rho _{2}^{\mathrm{Dim7}}(s,\alpha )=\frac{\langle \alpha _{s}G^{2}/\pi
\rangle \langle \overline{s}s\rangle }{288\pi ^{2}}\Theta (L_{2})\left[
m_{c}-m_{s}\alpha (\alpha -1)\right] ,\
\end{equation}%
and
\begin{equation}
\rho _{2}^{\mathrm{Dim8}}(s,\alpha )=\frac{\langle \overline{s}s\rangle
\langle \overline{s}g_{s}\sigma Gs\rangle }{24\pi ^{2}}\Theta (L_{2})\alpha
(\alpha -1).\
\end{equation}

Components of the function $\Pi (M^{2})$ are:%
\begin{eqnarray}
&&\Pi ^{\mathrm{Dim6}}(M^{2},\alpha ,\beta )=\frac{\langle
g_{s}^{3}G^{3}\rangle m_{s}}{15\cdot 2^{13}\pi ^{6}M^{2}\alpha ^{2}\beta
^{2}L^{3}N_{1}^{5}}\left\{ 2m_{c}M^{2}\alpha ^{4}\beta ^{4}L^{4}(3\beta
^{2}+\beta \alpha +\alpha ^{2})-\exp \left[ -\frac{m_{c}^{2}N_{2}}{%
M^{2}\alpha \beta L}\right] \right.  \notag \\
&&\times \left[ m_{c}^{5}N_{1}^{2}\left( 2\beta ^{10}+15\beta \alpha
^{7}(\alpha -1)^{2}+2\alpha ^{8}(\alpha -1)^{2}+\beta ^{9}(-4+15\alpha
)+\beta ^{5}\alpha ^{3}(-15+80\alpha -74\alpha ^{2})\right. \right.  \notag
\\
&&+2\beta ^{6}\alpha ^{2}(6+9\alpha -26\alpha ^{2})+\beta ^{7}\alpha
(15-32\alpha -3\alpha ^{2})-3\beta ^{3}\alpha ^{5}(5-6\alpha +\alpha
^{2})+4\beta ^{2}\alpha ^{6}(3-8\alpha +5\alpha ^{2})  \notag \\
&&\left. -4\beta ^{4}\alpha ^{4}(7-20\alpha +13\alpha ^{2})+\beta
^{8}(2-30\alpha +20\alpha ^{2})\right) -2m_{c}^{3}M^{2}\alpha ^{2}\beta
^{2}L^{2}\left( 2\beta ^{8}-5\beta \alpha ^{5}(\alpha -1)^{2}+2\alpha
^{6}(\alpha -1)^{2}\right.  \notag \\
&&-\beta ^{7}(4+5\alpha )+\beta ^{5}\alpha (-5+38\alpha -44\alpha
^{2})+\beta ^{6}(2+10\alpha -20\alpha ^{2})-22\beta ^{3}\alpha
^{3}(1-3\alpha +2\alpha ^{2})  \notag \\
&&\left. \left. \left. -2\beta ^{2}\alpha ^{4}(9-19\alpha +10\alpha
^{2})-2\beta ^{4}\alpha ^{2}(9-33\alpha +26\alpha ^{2})\right) \right]
\right\},
\end{eqnarray}%
\begin{eqnarray}
&&\Pi ^{\mathrm{Dim7}}(M^{2},\alpha ,\beta )=\frac{\langle \alpha
_{s}G^{2}/\pi \rangle \langle \overline{s}s\rangle m_{c}^{2}m_{s}}{72\pi
^{2}LN_{1}^{3}}\left[ \beta ^{4}+\alpha ^{2}(\alpha -1)^{2}+\beta
^{3}(-2+3\alpha )+\alpha \beta (2-5\alpha +\alpha ^{2})\right.  \notag \\
&&\left. +\alpha \beta (2-5\alpha +\alpha ^{2})+\beta ^{2}(1-5\alpha
+4\alpha ^{2})\right] ,
\end{eqnarray}

\begin{eqnarray}
&&\Pi ^{\mathrm{Dim8}}(M^{2},\alpha ,\beta )=-\frac{\langle \alpha
_{s}G^{2}/\pi \rangle ^{2}m_{c}}{81\cdot 2^{10}\pi ^{2}M^{6}\alpha \beta
(\beta -1)L^{6}N_{1}^{3}}\exp \left[ -\frac{m_{c}^{2}N_{2}}{M^{2}\alpha
\beta L}\right] \left\{ 24m_{c}^{4}m_{s}M^{2}(\beta -1)(\alpha +\beta
)^{3}L^{4}N_{1}^{2}\right.  \notag \\
&&+6m_{c}^{7}(\alpha +\beta )^{2}N_{1}^{5}-60m_{s}M^{6}\alpha \beta (\beta
-1)L^{5}\left[ \beta ^{4}-\beta ^{3}+\alpha ^{3}(\alpha -1)\right]
-9m_{c}M^{6}\alpha ^{2}\beta ^{2}(\beta -1)(\alpha ^{2}+\beta ^{2})L^{5}
\notag \\
&&-6m_{c}^{2}m_{s}M^{4}(\beta -1)L^{4}\left[ 8\beta ^{7}+18\beta \alpha
^{4}(\alpha -1)^{2}+8\alpha ^{5}(\alpha -1)^{2}+2\beta ^{6}(-8+9\alpha
)+7\beta ^{2}\alpha ^{3}(2-5\alpha +3\alpha ^{2})\right.  \notag \\
&&\left. +2\beta ^{3}\alpha ^{2}(7-15\alpha +8\alpha ^{2})+\beta ^{4}\alpha
(18-35\alpha +16\alpha ^{2})+\beta ^{5}(8-36\alpha +21\alpha ^{2})\right]
-2m_{c}^{5}M^{2}\left[ 12\beta ^{7}-3\beta ^{8}+3\alpha ^{4}\right.  \notag
\\
&&\times (\alpha -1)^{3}+\beta ^{6}(-18+6\alpha +\alpha ^{2})+3\beta \alpha
^{3}(\alpha -1)^{2}(-2+\alpha +3\alpha ^{2})+2\beta ^{5}(6-9\alpha +8\alpha
^{2})  \notag \\
&&+\beta ^{2}\alpha ^{2}(-6+14\alpha +11\alpha ^{2}-43\alpha ^{3}+24\alpha
^{4})+\beta ^{4}(-3+18\alpha -9\alpha ^{2}-30\alpha ^{3}+28\alpha
^{4})+\beta ^{3}\alpha \left( -6+14\alpha +6\alpha ^{2}\right.  \notag \\
&&\left. \left. -51\alpha ^{3}+37\alpha ^{4}\right) \right] \left[ \beta
^{4}+\alpha ^{2}(\alpha -1)^{2}+\beta ^{3}(-2+3\alpha )+\beta \alpha
(2-5\alpha +3\alpha ^{2})+\beta ^{2}(1-5\alpha +4\alpha ^{2})\right]
+m_{c}^{3}M^{4}\alpha \beta L^{2}  \notag \\
&&\times \left[ -3\beta ^{8}+3\alpha ^{4}(\alpha -1)^{3}+2\beta
^{7}(6+13\alpha )+6\beta ^{6}(-3-15\alpha +14\alpha ^{2})+\beta \alpha
^{3}(\alpha -1)^{2}(12-23\alpha +21\alpha ^{2})\right.  \notag \\
&&+\beta ^{5}(12+114\alpha -289\alpha ^{2}+177\alpha ^{3})+\beta ^{2}\alpha
^{2}(30-163\alpha +303\alpha ^{2}-244\alpha ^{3}+74\alpha ^{4})  \notag \\
&&\left. \left. +\beta ^{3}\alpha (12-181\alpha +467\alpha ^{2}-454\alpha
^{3}+156\alpha ^{4})+\beta ^{4}(-3-62\alpha +356\alpha ^{2}-493\alpha
^{3}+201\alpha ^{4})\right] \right\} ,
\end{eqnarray}

\begin{eqnarray}
&&\Pi ^{\mathrm{Dim9}}(M^{2},\alpha ,\beta )=\frac{\langle \alpha
_{s}G^{2}/\pi \rangle \langle \overline{s}g_{s}\sigma Gs\rangle
m_{c}N_{1}^{2}}{4608\pi ^{2}M^{4}\beta ^{4}\alpha ^{2}(\beta -1)^{3}L^{4}}%
\exp \left[ -\frac{m_{c}^{2}N_{2}}{M^{2}\alpha \beta L}\right] \left\{
48M^{4}\beta ^{3}(\beta -1)^{2}\alpha ^{2}(\alpha -1)L^{3}+6M^{4}\beta
^{3}\alpha \right.  \notag \\
&&\times (\beta -1)^{2}L^{3}+121M^{2}m_{c}m_{s}\beta ^{2}(\beta -1)\alpha
\left( \alpha -1\right) L^{2}N_{1}^{2}+40M^{2}m_{c}m_{s}\beta ^{2}(\beta
-1)L^{2}N_{1}^{2}-64M^{2}m_{c}m_{s}\beta \alpha ^{2}LN_{1}^{4}  \notag \\
&&\left. +32M^{2}m_{c}m_{s}\beta \alpha LN_{1}^{4}\left( 1+\alpha \right)
+16m_{c}^{3}m_{s}(\alpha +\beta )N_{1}^{5}\left( \alpha ^{2}-1\right)
+32m_{c}^{3}m_{s}\alpha (\alpha +\beta )N_{1}^{5}\right\} ,
\end{eqnarray}%
and
\begin{eqnarray}
&&\Pi ^{\mathrm{Dim10}}(M^{2},\alpha ,\beta )=-\frac{\langle \alpha
_{s}G^{2}/\pi \rangle \langle \overline{s}s\rangle ^{2}m_{c}^{2}N_{1}^{4}}{%
729\cdot 2^{5}M^{4}\pi ^{2}\alpha ^{2}\beta ^{4}(\beta -1)^{3}L^{4}}\exp %
\left[ -\frac{m_{c}^{2}N_{2}}{M^{2}\alpha \beta L}\right] \left\{
216m_{c}^{2}\pi ^{2}(\alpha -1)^{2}(\alpha +\beta )N_{1}^{3}-M^{2}\beta
L\right.  \notag \\
&&\times \left[ -g_{s}^{2}\alpha \beta (\alpha -1)(\beta -1)L+108\pi
^{2}\left( 4\beta ^{4}\alpha (\alpha -1)^{2}+4\alpha ^{3}(\alpha
-1)^{4}+\beta ^{3}(5-23\alpha +39\alpha ^{2}-24\alpha ^{3}+8\alpha
^{4})\right. \right.  \notag \\
&&\left. \left. \left. +\beta ^{2}(-10+39\alpha -69\alpha ^{2}+63\alpha
^{3}-40\alpha ^{4}+12\alpha ^{5})+\beta (5-20\alpha +38\alpha ^{2}-47\alpha
^{3}+48\alpha ^{4}-32\alpha ^{5}+8\alpha ^{6})\right) \right] \right\}.
\notag \\
&&
\end{eqnarray}

In expressions above, $\Theta (z)$ is Unit Step function. We have also used 
the following short-hand notations:
\begin{eqnarray}
N_{1} &=&\beta ^{2}+\beta (\alpha -1)+\alpha (\alpha -1),\ \ \ \ \
N_{2}=(\alpha +\beta )N_{1},\ L=\alpha +\beta -1,\   \notag \\
\ \ \ L_{1} &\equiv &L_{1}(s,\alpha ,\beta )=\frac{(1-\beta )}{N_{1}^{2}}%
\left[ m_{c}^{2}N_{2}-s\alpha \beta L\right] ,\ L_{2}\equiv L_{2}(s,\alpha
)=s\alpha (1-\alpha )-m_{c}^{2},\
\end{eqnarray}

\end{widetext}

\end{document}